\AtBeginDocument{%
  }
    
\documentclass[manuscript, screen]{acmart}

\DeclareMathDelimiter{(}{\mathopen} {operators}{"28}{largesymbols}{"00}
\DeclareMathDelimiter{)}{\mathclose}{operators}{"29}{largesymbols}{"01}

\usepackage{mathtools}  
\usepackage{paralist}

\usepackage[table]{xcolor}

\definecolor{lightblue}{RGB}{220,235,255}
\definecolor{lightred}{RGB}{255,220,220}

\newcommand{\tsd}{{$360^\circ$ }}

\usepackage{graphicx}

\usepackage[font=small]{caption}
\usepackage{subcaption}
\usepackage{bbm}

\usepackage{algorithm}
\usepackage{algpseudocode}

\definecolor{myred}{RGB}{220,43,25}
\definecolor{mygreen}{RGB}{0,139,0}
\definecolor{myblue}{RGB}{0,0,139}

\newcommand\mred[1]{{\mathcolor{myred}{\mathbf{#1}}}}
\newcommand\mblue[1]{{\mathcolor{myblue}{\mathbf{#1}}}}
\newcommand\mgreen[1]{{\mathcolor{mygreen}{\mathbf{#1}}}}

\usepackage{amsthm}

\theoremstyle{definition}

\DeclareMathOperator*{\argmin}{arg\,min}

\newcommand\nbUAVs{K}
\newcommand\nbChannels{C}

\newcommand\indicator{\mathbbm{1}}

\newcommand{\bb}[1]{{#1}}

\newcommand{\ie}{{\textit{i.e., }}}
\newcommand{\eg}{{\textit{e.g., }}}

\begin{document}

\title{ElasticVR: Elastic Task Computing in Multi-User Multi-Connectivity Wireless Virtual Reality (VR) Systems
}


\author{Babak Badnava}
\affiliation{%
  \institution{University of Kansas}
  \city{Lawrence, Kansas}
  \country{USA}}
\email{babak.badnava@ku.edu}

\author{Jacob Chakareski}
\affiliation{%
  \institution{New Jersey Institute of Technology}
  \city{Newark, New Jersey}
  \country{USA}}
\email{jacobcha@njit.edu}

\author{Morteza Hashemi}
\affiliation{%
  \institution{University of Kansas}
  \city{Lawrence, KS}
  \country{USA}}
\email{mhashemi@ku.edu}

\renewcommand{\shortauthors}{Badnava et al.}
\renewcommand{\shorttitle}{ElasticVR: Elastic Task Computing in Multi-User Multi-Connectivity Wireless Virtual Reality (VR) Systems}

\begin{abstract}

\bb{
Diverse emerging virtual reality (VR) applications integrate streaming of high fidelity 360$^\circ$ video content that requires ample amounts of computation and data rate.
Scalable 360$^\circ$ video tiling enables having \emph{elastic} VR computational tasks that can be scaled adaptively in computation and data rate based on the available user and system resources.
}
We integrate scalable 360$^\circ$ video tiling in an edge-client wireless multi-connectivity architecture for joint elastic task computation offloading across multiple VR users called \emph{ElasticVR}.
To balance the trade-offs in communication, computation, energy consumption, and quality of experience (QoE) that arise herein, 
we formulate a constrained QoE and energy optimization problem that integrates the multi-user/multi-connectivity action space with the elasticity of VR computational tasks.
The ElasticVR framework introduces two multi-agent deep reinforcement learning solutions, namely the centralized phasic policy gradient (CPPG) and the independent phasic policy gradient (IPPG).
CPPG adopts a centralized training and centralized execution approach to capture the coupling between users’ communication and computational demands.
This leads to globally coordinated decisions at the cost of increased computational overheads and limited scalability.
To address the latter \bb{challenges}, we also explore an alternative strategy denoted IPPG that \bb{adopts} a centralized training with decentralized execution paradigm.
IPPG leverages shared information and parameter sharing to learn robust policies; however, during execution, each user takes action independently based on its local state information only. 
The decentralized execution alleviates the communication and computation overhead of centralized decision-making and improves scalability.
We train our methods with real-world 4G, 5G, and WiGig network traces and $360^\circ$ video datasets to evaluate their performance in terms of delivered video quality, response time, energy consumption, and scalability.
\bb{
In particular, CPPG reduces the latency by 28\% and energy consumption by 78\% compared to IPPG in an environment with three VR users.
However, IPPG outperforms CPPG as the number of users increases by benefiting from a smaller state space and parameter sharing, which lead to a higher sample efficiency.
}
We show that IPPG sustains strong performance across diverse workloads, achieving a 19.54\% improvement in PSNR, 20\% reduction in response time, and 35.13\% reduction in energy consumption compared to CPPG.
\bb{
Moreover, we show that the ElasticVR framework improves the PSNR by 43.21\%, while reducing the response time and energy consumption by 42.35\% and 56.83\%, respectively, compared with a case where no elasticity is incorporated into VR computations.
}

\end{abstract}



\maketitle

\section{Introduction}\label{sec:Intro}
It is envisioned that next generation wireless networks (6G and beyond) will enable an unprecedented proliferation of computationally-intensive and bandwidth-hungry applications such as augmented/virtual reality (AR/VR) and online 3D gaming~\cite{SeaGate-2019-State}.
The VR use cases 
hold tremendous potential to advance our society and impact our daily life and the economy~\cite{ChakareskiKY:20,Chakareski-2023-Millimeter}.
Currently, VR applications are becoming increasingly popular in education, training, healthcare, and gaming,
reaching a global market size of US \$20.83 billion in 2025~\cite{fortune-2024-vr}.
These emerging VR applications require streaming of high fidelity $360^\circ$ video content, which requires an ample amount of computational and communication resources near the edge of the network.
For instance, a minimum of $12$K high-quality spatial resolution and $100$ frames per second (FPS) temporal rate are recommended by MPEG for $360^\circ$ VR~\cite{Chakareski-2020-6DOF}.
\bb{
In particular, VR applications need to decode and render virtual scenes in real-time.
The pipeline is composed of computationally heavy tasks such as 360$^\circ$ video encoding, decoding, spatial projection, stitching, and rendering~\cite{Chakareski-2020-6DOF}.
}
Furthermore, VR devices have limited on-board energy resources.
High computational demands, continuous sensor input, and wireless connectivity drain batteries quickly, limiting the usage time of VR devices. Hence, VR \bb{computation and communication resource allocation methods} must be designed with energy-efficiency in mind.


\bb{Traditional VR computational tasks exhibit a binary nature, meaning they require a fixed amount of computational and communication resources with a non-negotiable deadline.
However, the scalable 360$^\circ$ video tiling~\cite{Chakareski-2020-6DOF, Badnava-neural-2025} enables adaptive modulation of rendering quality, frame rates, and latency-sensitive computations. 
This adaptive modulation introduces \emph{elastic} VR computation, meaning tasks can be dynamically adjusted based on various user requirements as well as system capabilities and constraints.
This elasticity, in turn, provides more flexibility for an efficient communication and computation resource management to provide an enhanced QoE.

}

Within this context, our goal is to model and integrate \emph{elasticity} of VR computation to enable 
efficient resource management (computing and communication), coordination among distributed VR users, and guaranteed QoE.
To this end, we consider several requirements \bb{and constraints}:
\bb{(i) \emph{user-imposed} requirements that determine the computation deadline and QoE,}
(ii) \emph{network-imposed} constraints that determine the available communication link rate,
(iii) \emph{computation-imposed} constraints that determine the available computational resources for  VR tasks,
and (iv) \emph{device-imposed} constraints that determine the available computation and energy resources on VR devices.

\begin{figure}[t]
    \centering
    \includegraphics[width=.95\linewidth, trim= 2mm 2mm 1mm 70mm, clip=true]{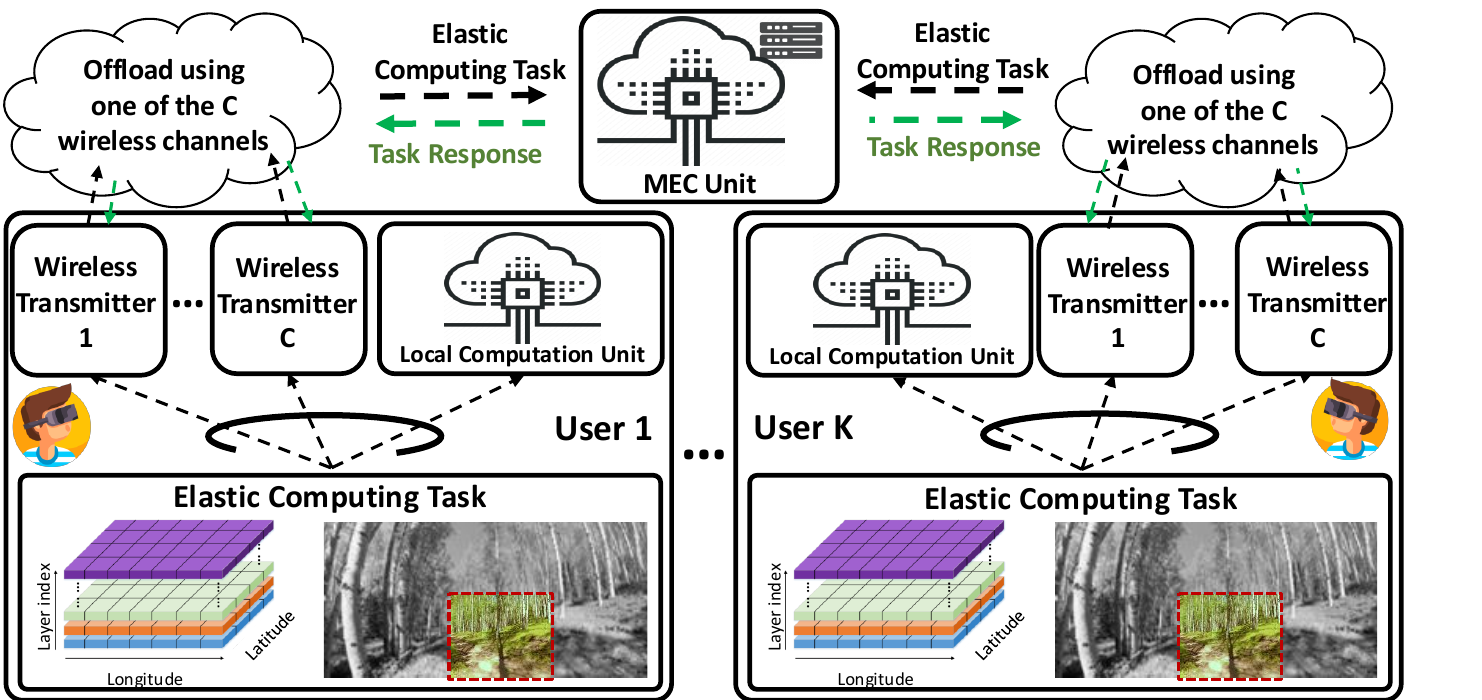}
    \caption{Multi-user multi-connectivity edge-assisted VR offloading scheme: VR users can compute their elastic tasks locally or offload them to a nearby MEC unit through one of the $\nbChannels$ wireless technologies available. Wireless technologies operate on different bands with unique characteristics.}
    \label{fig:sys-model}
\end{figure}

In this paper, we develop a novel constrained QoE and energy optimization problem for elastic VR task offloading in a multi-user multi-connectivity VR system, as illustrated in Fig. \ref{fig:sys-model}.
Here, elastic computational tasks can be decoded and/or rendered by a multi-access edge computing (MEC) unit, located on the premises of the VR arena.
On one hand, the MEC provides more computational resources, thus can process the elastic computational tasks at a higher rate, which leads to a lower latency and better QoE for users.
On the other hand, computing tasks on the MEC introduce a higher communication energy consumption and a higher bandwidth requirement since elastic tasks need to be transmitted over the air, which leads to a higher communication latency.
Moreover, elastic tasks can be adjusted, via an \emph{elasticity parameter}, to account for varying system conditions such as time-varying network conditions.
Our proposed decision-making framework considers the interplay between the variable communication and computation requirements of elastic VR tasks, as well as limited on-board energy constraints.
Leveraging state-of-the-art multi-agent reinforcement learning (MARL) paradigms, we present two MARL agent architectures, called CPPG and IPPG.
They solve a constrained multi-user multi-connectivity QoE and energy optimization problem, while incorporating user performance requirements (\ie task deadlines).
These decision-making agents learn the optimal offloading action and elasticity parameter in a VR arena by considering the communication history (\ie past average uplink and downlink throughput) and elastic task information (\ie task size and computational intensity).
In summary, our main contributions are as follows:
\begin{enumerate}
    \item \emph{Constrained QoE and energy optimization problem.}
    We introduce a multi-user, multi-connectivity edge-assisted VR offloading scheme, in which VR users can offload their elastic tasks to a nearby MEC unit. We then formulate a constrained QoE and energy optimization problem to determine the optimal offloading policy with respect to QoE constraints, network conditions, and the computational and communication requirements of elastic tasks.
    
    \item \emph{Multi-agent elastic task offloading framework and simulator.}
    We develop a multi-agent learning-based offloading framework that introduces two architectures, CPPG and IPPG, to learn optimal offloading policies that maximize the energy-efficiency while satisfying QoE requirements (\ie task deadline).
    We also develop a trace-driven Gym-like~\cite{brockman2016openai} VR task offloading simulator using a dataset of real-world VR tasks, VR user head movement and mobility information, and 4G, 5G, and WiGig network traces.
    The simulator and framework are open-sourced and available on \href{https://github.com/badnava-babak/vr-streaming}{Github}.
    
    \item \emph{Extensive numerical analysis.}
    Leveraging the developed simulator, we conduct an extensive numerical analysis under divers system conditions.
    \bb{First, we characterize the optimal feasible solution set by providing the full range of trade-offs among the three competing objectives (\ie response time, video quality, energy consumption) and their Pareto frontiers.
    The Pareto frontiers reveal the boundary of solutions that cannot be further improved in one objective without sacrificing at least of the others.
    We show that our framework provides a close-to-optimal performance relative to these boundaries.
    Furthermore, our results demonstrate inherent trade-offs between centralized and decentralized decision-making processes in terms of scalability.
    In a scenario with three VR users, CPPG reduces the latency by 28\% and energy consumption by 78\% compared to the decentralized approach (\ie IPPG).
    However as the number of users increases, the decentralized approach (\ie IPPG) outperforms the centralized approach (\ie CPPG) by benefiting from a smaller state space and parameter sharing, which leads to a higher sample efficiency.
    Moreover, we show that introducing computation \emph{elasticity} leads to substantial gains compared with an elasticity-agnostic baseline.
    In particular, the introduced \emph{elasticity} increases the PSNR by 43.21\% while reducing the response time and energy consumption by 42.35\% and 56.83\%, respectively.
    Finally, we analyze the effect of heterogeneous content on the ElasticVR framework. 
    Our analysis shows that IPPG provides a 19.54\% improvement in PSNR, a 20\% reduction in response time, and a 35.13\% reduction in energy consumption compared to CPPG.
    }
    
    
\end{enumerate}
The rest of this paper is organized as follows. 
In Section \ref{sec:related}, we review related works and highlight our contributions.
In Section \ref{sec:system-model}, we present the edge-assisted VR elastic task offloading system model.
In Section \ref{sec:problem}, we present the constrained QoE and energy optimization problem, the challenges involved in solving such a problem, and a discussion on the optimal solution set.
In Section \ref{sec:solution}, we present our learning-based multi-agent decision making framework.
In Section \ref{sec:evaluation}, we provide simulation results and performance analysis of the proposed architectures. Section \ref{sec:conclusion} concludes the paper.

\section{Related Works}\label{sec:related}
\textbf{Generic Task Offloading:}
MEC has emerged as a promising solution that provides extra computational resources near the edge of the network for latency-sensitive and resource-hungry applications.
However, several technical challenges need to be considered and optimized in order to have a reliable task offloading framework.
A plethora of prior studies have investigated various aspects of task offloading services in a general setting.
A group of works investigated the problem of latency minimization~\cite{Ouyang-2019-Adaptive, Molina-2014-Joint, Wu-2021-EdgeCentric, Wang-2022-Decentralized, Yang-2022-Optimal, Huang-2019-Fine, Jia-2021-Learning}.
The works presented in \cite{Liu-2020-Joint, Zhu-2018-Cooperative, Cao-2020-UAV, Zhang-2018-Energy} investigated the energy-efficiency aspects of the task offloading services.
The authors in \cite{Jiang-2023-Joint} formulated an optimization problem that provides a guaranteed QoE while meeting long-term MEC energy constraints.
Another group \cite{Yang-2022-Multi, Sacco-2021-Sustainable, Sacco-2022-Self} investigated the trade-off between latency and energy consumption in task offloading services.
We advance these studies by accounting for unique aspects of VR applications such as task elasticity, shared computational resources at the edge, and QoE considerations.





{\bf MEC-Assisted VR Systems.} Integrating MEC in VR systems has been explored to provide additional: (i) computational power for decoding, rendering, and stitching of $360^\circ$ videos \cite{Chakareski-2020-6DOF, Hsu-2020-MEC} and (ii) storage space for caching $360^\circ$ video content~\cite{Dai-2020-View, Maniotis-2021-Tile,Dang-2019-Joint}.
Moreover, MEC node placement \cite{Zhang-2022-UAV,Liu-2023-On}, MEC architecture for on-demand/real-time streaming \cite{Guo-2021-Design, Aung-2024-Edge,Han-2019-Real,Zhang-2023-RealVR}, communication resource management \cite{Yang-2018-Communication, Lin-2021-Resource, Guo-2020-AnAW, Chen-2019-Data}, and user scheduling \cite{Huang-2018-MAC} have been explored. Finally, MEC integrated with high-bandwidth mmWave links and multi-connectivity \cite{Chakareski-2024-Live, Ge-2017-Multipath, Liu-2019-MEC, Gupta-2023-mmWave, Ren-2019-Edge, Chakareski-2023-Millimeter} has been studied to enable multi-Gbps data rates needed for lifelike VR immersion. 
We advance these studies by formulating a dynamic decision-making framework for adaptive elastic task offloading in a multi-connectivity VR system.
Our framework establishes a \emph{user-edge computing continuum}, which dynamically makes offloading decisions for elastic tasks under time-varying wireless network conditions and computational requirements. This, in turn, enhances the flexibility of the system to further improve the energy-efficiency and latency by taking into account the communication, computation, and energy requirements of elastic tasks.

\textbf{VR Computation Offloading and Task Elasticity:}
Among the works that integrated MEC into VR application, a multitude of works have investigated the non-elastic task offloading problem for VR applications to reduce the latency or improve the energy-efficiency.
The authors in \cite{Younis-2020-Latency} proposed a three-layered architecture \bb{involving the end user, the mobile edge, and the cloud to find an efficient application placement on the MEC-AR layers to minimize the network latency.
}
The authors in \cite{Song-2023-Computing} used deep reinforcement learning for offloading decision-making to reduce computational latency.
The authors in \cite{Cheng-2022-Design} propose a framework in which the VR computational tasks are partially offloaded to the MEC to meet the required latency for VR video streaming applications.
To tackle the energy consumption problem, the authors in \cite{Didar-2023-eAR} considered computation and device-imposed constraints by proposing a framework for mobile AR computing to reduce energy and storage consumption.
Moreover, the authors in \cite{Wang-2023-LEAF} proposed an edge-based energy-aware AR system that enables AR devices to dynamically change their configurations
based on user preferences
to improve the energy-efficiency of an object detection task.
The authors in \cite{Nyamtiga-2022-Edge} investigated the benefits of task offloading in terms of computational load and power consumption reduction. 
The authors in \cite{Ren-2022-adative} introduced an offloading framework that adaptively downscales a mobile AR video feed to reduce upload latency, then uses super-resolution at the edge server to restore quality.
This work focuses on single-user 4K VR streaming over Wi-Fi, dynamically trading off resolution for latency, which parallels the user’s ``elastic'' task concept (adapting task fidelity to network limits).
While these studies focus on alleviating the computation and device-imposed constraints, 
\bb{we leverage an elastic task model (\ie scalable 360$^\circ$ video model) and design a joint communication and computation resource management algorithm to balance
}
the computational and communication trade-offs \bb{in VR systems}. 
Moreover, our work leverages multi-connectivity technology to further balance the communication, computation, energy, and QoE trade-offs in networked VR systems.

\begin{table}[t]
\centering
\caption{Notation Table}
{
\begin{tabular}{ll}
\hline
\textbf{Symbol} & \textbf{Description} \\
\hline
$k \in \{1,\dots,K\}$ & User index \\
$c \in \{1,\dots,C\}$ & Communication channel index \\
$u_k \in \{0,1,\dots,C\}$ & Offloading decision of user $k$ ($0$: local, $c$: channel $c$) \\
$e_k \in \{1,\dots,L\}$ & Elasticity parameter of user $k$ \\
$S_k(e_k)$ & Task size (bits) \\
$I_k(e_k)$ & Task computational intensity (cycles/bit)\\
$S^{\text{res}}(e_k)$ & Task result size (bits) \\
$T_k^d$ & Task deadline (seconds) for user $k$ \\
$T_k^r(e_k,u_k)$ & Task response time \\
$T^c(S_k(e_k),I_k(e_k),f)$ & Task computation time \\
$T_k^{\text{tx}}(S_k(e_k),u_k)$ & Task transmission time\\
$T_k^{\text{rx}}(S^{res}_k(e_k),u_k)$ & Task result reception time\\
$R_k(u_k)$ & Average transmission rate of user $k$ on channel $u$ \\
$f_k^{\text{vr}}$ & CPU frequency of VR headset \\
$f_k^{\text{mec}}$ & CPU frequency allocated by MEC to user $k$ \\
$z(f)$ & Computation speed function at CPU frequency $f$ \\
$E_k^{\text{tot}}(e_k,u_k)$ & Total energy consumption \\
$E_k^c(S_k(e_k),I_k(e_k),f_k^{vr})$ & Local computation energy consumption \\
$E_k^{\text{tx}}(S_k(e_k),u_k)$ & Transmission energy consumption \\
$E_k^{\text{rx}}(S_k(e_k),u_k)$ & Reception energy consumption \\
$P_k^{\text{tx}}(u_k)$ & Transmission power profile \\
$P_k^{\text{rx}}(u_k)$ & Reception power profile \\
\hline
\end{tabular}
}
\end{table}

\section{Edge-Assisted VR Task Offloading Modeling}\label{sec:system-model}
As depicted in Fig. \ref{fig:sys-model}, we consider a multi-connectivity MEC network that consists of $\nbUAVs$ VR users.
All VR headsets are equipped with computational resources (\ie CPU) to perform computational tasks arriving at the VR headsets.
Upon arrival of each elastic task (\ie \tsd video decoding), a decision-making agent \bb{makes a joint decision on the elasticity parameter $e_k$ and offloading action $u_k$.
The offloading decision determines whether} to perform the task locally or offload it to the MEC unit using one of the $C$ available communication channels. Then, the elastic task will be computed (on the VR headset or the MEC unit) and the result will be sent to the VR user.

\begin{figure}[t]
    \centering
    \includegraphics[width=.95\linewidth, trim={4.3cm 5.3cm 4.3cm 3.8cm},clip]{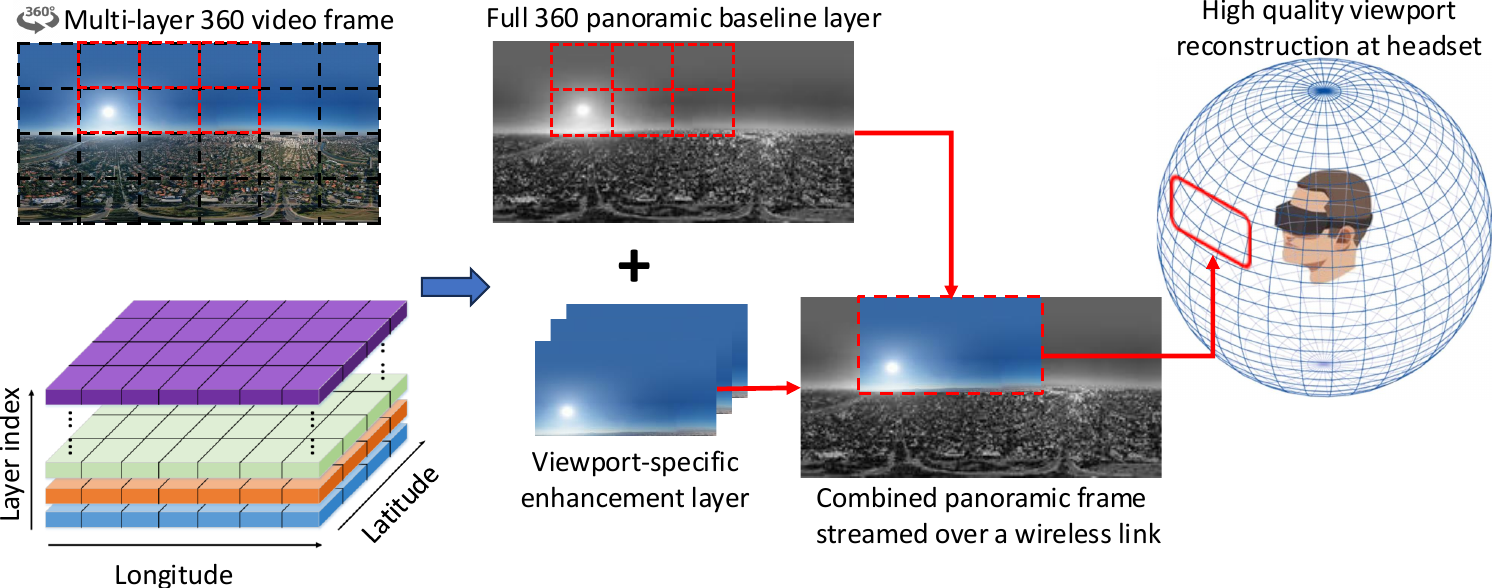}
    \caption{Multi-layer $360^\circ$ video model (lower left corner): Viewport-specific layers combined with a wide $360^\circ$ panorama layer are transmitted to a VR headset. As the number of added layers increases, the video bitrate and the video quality increase, which leads to a higher QoE. However, the required computational and communication resources increase, which leads to a higher latency and energy consumption, thus a lower QoE.}
    \label{fig:video-model}
\end{figure}

\subsection{VR System Computation Modeling}\label{sec:comp-model}


\textbf{Elastic Task Model:}
We consider a sequence of elastic tasks (\ie decoding and rendering of a group of pictures (GoP)) arriving at VR user $k$.
Each of these tasks is characterized by four different features: 
\begin{inparaenum}[(i)]
    \item task size/length (in bits),  denoted by $S_k(e_k)$;
    \item task computational intensity (in CPU cycles per bit), denoted by $I_k(e_k)$;
    \item task deadline (in seconds), denoted by $T_k^d$; and
    \item task response size/length (in bits), denoted by $T_k^{res}(e_k)$. 
\end{inparaenum}
These tasks can adapt their resource consumption or execution behavior in response to changing system conditions by changing a parameter $e_k \in \{1, .., L\}$, which we call the \emph{elasticity parameter}.
A larger $e_k$ leads to a higher QoE for users at the expense of larger task size and computational intensity, which leads to higher computation and communication resource requirements.
An example of such an elastic task is the scalable $360^\circ$ video tiling design~\cite{Chakareski-2020-6DOF, Badnava-neural-2025}.
\bb{
As depicted in Fig.~\ref{fig:video-model}, each panoramic $360^\circ$ video frame is partitioned into tiles arranged in a $H \times V$ grid.
Some of these tiles are located within the user's current viewport.
We employ user head movement navigation information to determine which tiles are located within the user's viewport.
There are $L$ layers of increased immersion fidelity for each tile in the panoramic $360^\circ$ video frame.
Each layer incrementally increases the transmission video bitrate and, subsequently, the video quality delivered to the user.
A block of consecutive video frames, compressed together with no reference to others, creates a GoP or video segment.
For each GoP, we construct a baseline representation of the entire $360^\circ$ panorama by using the base layer for all tiles.
Then, we construct an enhancement representation by integrating the enhancement layers for the tiles within the user's current viewport.
Let $b_{h,v}(l)$ denote the size, in bits, of the tile on row $h$, column $v$, and layer $l$. 
Then, the size of the GoP (\ie task size), denoted by $S_k(e_k)$, is determined by summing over all tiles' sizes in the GoP. Thus, we have: 
\begin{equation}
S_k(e_k) := 
\begin{cases}
    e_k =0 \quad \Rightarrow & \quad \sum_{\forall h, v} b_{h, v}(0) \quad  \\
    e_k > 0 \quad \Rightarrow & \quad S_k(0) + \sum_{l=1}^{e_k} \sum_{\forall h,v} m_{h,v}b_{h, v}(l).
\end{cases}
\end{equation}
Here, $m_{v,h} \in \{0, 1\}$ is a binary value denoting whether the tile in row $h$ and column $v$ is located in user's viewport or not given the user's head navigation information (yaw, pitch, and roll) and the user's horizontal and vertical field of view (FOV).
Similar scalable 360$^\circ$ tiling and viewport-adaptive layered representations have been used in prior work on 360$^\circ$ video streaming~\cite{Chakareski-2020-Viewport}.
Moreover, we assume a linear relation between task size and computational intensity.
Hence, the computational intensity of each task is defined by $I_k(e_k) = \beta S_k(e_k)$, where $\beta > 0$.

}


Upon arrival of each task (\ie decoding and rendering of a GoP) at the $k^{th}$ user, a decision-making agent \bb{makes a joint decision on the elasticity parameter $e_k$ and offloading action $u_k$.
The offloading decision determines} whether the task would be computed on the VR headset or offloaded to the MEC unit for computation using one of the $C$ communication channels. 
We denote $u_k \in \{0, 1, \dots C\}$ as the \bb{offloading} decision, which is defined as follows:
\begin{equation}
\label{eq:action}
    u_k := 
    \begin{cases}
         u_k = 0 \;\; \Rightarrow & \text{Compute on VR headset.} \\
        u_k > 0 \;\; \Rightarrow & \text{Offload via channel $u_k$.}
    \end{cases}
\end{equation}
In case of local computation (\ie $u_k = 0$), the task is computed on the local computing unit 
of the $k^{th}$ user.
Alternatively, when the decision-making agent decides to offload the task (\ie $u_k > 0$), the task is transmitted to the MEC unit through the communication channel with index $u_k$.
Then, on the MEC side, the task is computed by the MEC unit, and the computation result is returned to the user through the same communication channel.
This leads to the following task response time:
\begin{equation}
    T_k^r(e_k, u_k) =
        \begin{cases}
                u_k = 0 \;\; \Rightarrow & T^c\Big(S_k(e_k), I_k(e_k), f_k^{vr}\Big) , \\
                u_k > 0 \;\; \Rightarrow & T^{tx}_k\Big(S_k(e_k), u_k\Big)  +  T^c\Big(S_k(e_k), I_k(e_k), f^{mec}_k\Big) +  T_k^{rx}\Big(S^{res}_k(e_k), u_k\Big). 
        \end{cases}
\end{equation}
Here, the function $T^c\Big(S_k(e_k), I_k(e_k), f_k^{vr}\Big)$ returns the task computation time given the elastic task size $S_k(e_k)$, computational intensity $I_k(e_k)$, and VR headset's CPU frequency $f_k^{vr}$. 
The function $T^{tx}_k\Big(S_k(e_k), u_k\Big)$ returns the transmission time and $T_k^{rx}\Big(S^{res}_k(e_k), u_k\Big)$ returns the task result transmission time.
Moreover, $f_k^{mec}$ denotes CPU frequency that the MEC unit has allocated to the $k^{th}$ user.


\textbf{Elastic Task Computation Model:}
The elastic task computation time depends on its size, computational intensity and the operating frequency of the CPU.
Given a CPU with an operating frequency of $f$ (in cycles per second), an elastic task with size $S_k(e_k)$ (in bits) and computational intensity $I_k(e_k)$ (in cycles per bit), the task computation time is calculated as:
\begin{align}\label{eq:cpu-exec-model}
    T^c \Big(S_k(e_k), I_k(e_k), f\Big) = \frac{S_k(e_k)I_k(e_k)}{f} \quad [Seconds].
\end{align}

\textbf{VR System Communication Model:}
The VR device transmits the task through one of the $C$ available channels to the MEC server.
The expected transmission rate for a task is modeled as:
\begin{equation}
    R_k (u_k) = \frac{1}{t_e - t_s} \int_{t_s}^{t_e} R_k^t(u_k) dt,
\end{equation}
where $t_s$ and $t_e$ are the transmission start and end times, respectively, and $R_k^t(u_k)$ is the throughput provided by the uplink communication channel with index $u_k$ for the $k^{th}$ user.
The expected reception rate for a task result is modeled in a similar way over the downlink communication channel. 
Hence, the task transmission time and result reception time follow:
\begin{equation}\label{eq:tx-time}
    T^{tx}_k\Big(S_k(e_k), u_k\Big) = \frac{S_k(e_k)}{R_k(u_k)}, \;
    T_k^{rx}\Big(S^{res}_k(e_k), u_k\Big) = \frac{S^{res}_k(e_k)}{R_k (u_k)}.
\end{equation}
Note that although we use the same notation for uplink and downlink throughput, we employ different traces for each of them in our simulations and numerical evaluations.

\subsection{Energy Consumption Modeling}
The VR headset's battery is depleted by three different components: its CPU for local computation, its transmitter for offloading tasks to the MEC unit, and its wireless receiver for receiving results of computations.

\textbf{Computation Energy Consumption:}
The main source of task computation energy consumption is the CPU, and the energy consumption of the other components are negligible~\cite{Zhang-2013-Energy}.
CPU power usage consists of different factors such as dynamic power, short circuit power, and leakage power. Among all these three components, the dynamic power is the dominant part~\cite{Zhang-2013-Energy, Burd-19960-Processor}.
The dynamic power consumption is proportional to the squared supplied voltage to the chip~\cite{Burd-19960-Processor}. 
Furthermore, the clock frequency of the CPU is approximately linearly proportional to the supplied voltage~\cite{Burd-19960-Processor}. 
Thus, for user $k$, the headset's CPU  energy consumption follows:
\begin{align}
    E_k^c \Big(S_k(e_k), I_k(e_k), f_k^{vr}\Big) = \kappa S_k(e_k) I_k(e_k) (f_k^{vr})^2 \quad [Joule],
\end{align}
where $\kappa$ is the CPU capacitance factor, specific to each CPU, and $f_k^{vr}$ is the VR headset's CPU frequency.
$S_k(e_k)$ and $I_k(e_k)$ are elastic task size and computational intensity of the $k^{th}$ user, respectively.
Note that we do not consider the elastic task computation energy consumption for the offloaded tasks since the MEC unit has unlimited power supplies.

\textbf{Communication Energy Consumption:}
The power consumption for transmission and reception of a task over a wireless channel varies based on the wireless technology used by the headset (i.e. Wi-Fi, WiGig, LTE, or 5G). 
The wireless transmitter energy consumption can be summarized by three different factors: transmission power, the efficiency of the transmit power amplifier, and a constant factor that corresponds to the power consumed by all other circuit blocks~\cite{Dusza-2013-CoPoMo, Zappone-2015-Energy}. 
The transmit power amplifier is usually assumed to be 100\% efficient, and the constant power consumption factor is negligible~\cite{Dusza-2013-CoPoMo, Zappone-2015-Energy}. 
Hence, the transmitter energy consumption is modeled as:
\begin{align}\label{eq:tx-energy-consum}
    E_k^{tx}\Big(S_k(e_k), u_k\Big) = T^{tx}_k\Big(S_k(e_k), u_k\Big) P^{tx}_k(u_k) \quad [Joule],
\end{align}
where the function $T^{tx}_k\Big(S_k(e_k), u_k\Big)$ returns the task transmission time over the wireless channel with index $u_k$ as described in Eq. \eqref{eq:tx-time}, and $P^{tx}_k(u_k)$ is the power allocation profile for the wireless channel with index $u_k$.
Similarly, the receiver energy consumption is calculated as:
\begin{align}\label{eq:rx-energy-consum}
    E_k^{rx}\Big(S_k(e_k), u_k\Big) = T^{rx}_k\Big(S^{res}_k(e_k), u_k\Big) P^{rx}_k(u_k) \quad [Joule],
\end{align}
where $T^{rx}_k\Big(S^{res}_k(e_k), u_k\Big)$ denotes the task result transmission time over wireless channel $u_k$, and $P^{rx}_k(u_k)$ is the power consumption profile of the wireless receiver for wireless channel $u_k$.

Note that the total task energy consumption is calculated differently for locally computed tasks and offloaded tasks.
The energy consumption of locally computed tasks consists of only the headset's CPU energy consumption,
while the energy consumption of offloaded tasks consists of transmission and reception energy consumption. 
Hence, the total energy consumption of an elastic task is obtained by:
\begin{equation}
    E^{tot}_k(e_k, u_k) =
        \begin{cases}
            u_k = 0 \;\; \Rightarrow & \kappa S_k(e_k) I_k(e_k) (f_k^{vr})^2 \quad  \\
            u_k > 0 \;\; \Rightarrow & T^{tx}_k\Big(S_k(e_k), u_k\Big) P^{tx}_k(u_k) +  T^{rx}_k\Big(S^{res}_k(e_k), u_k\Big) P_k^{rx}(u_k).
        \end{cases} \\
\end{equation}

\section{Constrained QoE and Energy Optimization}\label{sec:problem}
In this section, we formulate a QoE and energy optimization problem for multi-user multi-connectivity immersive computing systems with elastic tasks. 
We first examine the factors that impact the QoE \bb{and energy consumption} in such a setting.
One of the key goals for VR systems is to improve the QoE that can lead to increased long-term user engagement.
While users may differ in their preference, there are three key contributing factors to the QoE \bb{and long-term user engagement:} 
\begin{enumerate}
    \item \textit{Video Quality:} The average video quality \bb{perceived by users} $Q(\mathbf{e})= \frac{1}{K} \sum_{k=1}^{K} q(e_k)$. 
    \item \textit{Task Response Time:} The average task response time, $T(\mathbf{e}, \mathbf{u}) = \frac{1}{K} \sum_{k=1}^{K} T_k^r(e_k, u_k)$, affects the system performance and QoE negatively, thus needs to be minimized.
    \item \textit{Energy Consumption:} A lower average energy consumption, $E(\mathbf{e}, \mathbf{u}) = \frac{1}{K} \sum_{k=1}^{K} E^{tot}_k(e_k, u_k)$, can lead to longer service time, \bb{thus long-term user engagement.}
\end{enumerate}
Here $\mathbf{e} = \big[e_1, e_2, ..., e_K\big]$ and $\mathbf{u} = \big[u_1, u_2, ..., u_K\big]$, respectively, denote the elasticity parameter and offloading decisions made for all users for the \bb{current computation task.} We consider a weighted sum of these factors to make sure that all key contributing factors are considered.
Therefore, the optimization objective is formulated as:
\begin{align}\label{eq:qte}
    \mathit{QTE}(\mathbf{e}, \mathbf{u}) = w_0 Q(\mathbf{e}) - w_1 T(\mathbf{e}, \mathbf{u}) - w_2 E(\mathbf{e}, \mathbf{u}).
\end{align}

While there are various choices for the function $q(\cdot)$, we use the PSNR of the luminance component of the 360$^\circ$ video signal over the viewer's FoV, which is commonly denoted as Y-PSNR.
\bb{
PSNR is computed using the distortion of each tile in the user's viewport.
Let $y_{h,v}(e_k)$ denote the mean squared error (MSE) of the luminance (Y) component of the tile in row $h$, column $v$, and layers one up to $e_k$ of the GoP. 
Then, the average PSNR value over the user's current viewport is given by:
\begin{align}
    q(e_k) 
    = \frac{1}{\sum_{\forall h,v} m_{h,v}} 
       \sum_{\forall h,v} m_{h,v} 
       \left[ 10 \log_{10}\!\left( \frac{255^2}{y_{h,v}(e_k)} \right) \right], 
\end{align}
where, $m_{h,v}$ denotes whether the tile in row $h$ and column $v$ is within the user's viewport or not, as described in Section \ref{sec:comp-model}.
}
PSNR is a widely used metric for measuring the quality of the video perceived by end users in VR applications.
Its widespread use as an objective video-quality metric is due to several factors.
PSNR is easy to calculate, has a clear physical meaning, and is mathematically convenient for optimization~\cite{Zhou-2004-SSIM}.



\subsection{Constrained QoE and Energy Optimization Problem Formulation}
Our constrained QoE and energy optimization problem has two sets of decision variables.
In particular, $e_k \in \{1, 2, .., L\}$ determines the elasticity parameter, and $u_k \in \{0, 1, \dots C\}$ determines the offloading decision according to Eq. \ref{eq:action}.
In addition to these decision variables, MEC computational resources need to be allocated to the users, which we assume that they have been allocated proportional to user's requirements \bb{and calculated by employing the computational intensity of offloaded tasks as: $\frac{I_k(e_k)}{\sum_{k=1}^K \indicator_{\{u_k > 0\}} I_k(e_k)}$.}
Therefore, we formulate the optimization problem outlined in Eq. \eqref{eq:ee-maximization}:
\begin{subequations}\label{eq:ee-maximization}
\begin{alignat}{5}
    \max_{\mathbf{e}, \mathbf{u}} \;\; & \mathit{QTE}(\mathbf{e}, \mathbf{u}) \quad \tag{\ref{eq:ee-maximization}} \\
        \text{s.t.} \;\; & 
        T_k^r(e_k, u_k) \leq T_k^d \quad \forall \; k \in \{1,2,...,K\} \label{eq:deadline-const}, \\
        &
            T_k^r(e_k, u_k) 
                 = \indicator_{\{u_k = 0\}} 
                    \Big[ T^c\Big(S_k(e_k), I_k(e_k), f_k^{vr}\Big) \Big] + \indicator_{\{u_k > 0\}} 
                \Big[
                    T^{tx}_k\Big(S_k(e_k), u_k\Big) 
                     +  T^c\Big(S_k(e_k), I_k(e_k), f^{mec}_k\Big)
                     +  T^{rx}_k\Big(S^{res}_k(e_k), u_k\Big) 
                \Big]
        \label{eq:tr-const}, \\
        \quad &
            E^{tot}_k(e_k, u_k)
             = \indicator_{\{u_k = 0\}}
                \Big[ \kappa S_k(e_k) I_k(e_k) (f_k^{vr})^2 \Big]  
            + \indicator_{\{u_k > 0\}} 
                \Big[ 
                    T^{tx}_k\Big(S_k(e_k), u_k\Big) P^{tx}_k(u_k) 
                     + T^{rx}_k\Big(S^{res}_k(e_k), u_k\Big) P_k^{rx}(u_k) 
                \Big]
        \label{eq:energy-const}, \\
        \;\; & \sum_{k=1}^K \indicator_{\{u_k > 0\}} z(f_k^{mec}) \le Z_{mec}. \label{eq:mec-const}
\end{alignat}
\end{subequations}
\hrulefill

\noindent
Here, Eq. \eqref{eq:deadline-const} ensures that all users meet their task deadlines.
Eq. \eqref{eq:tr-const} defines the computation system model for each elastic task.
Eq. \eqref{eq:energy-const} defines the energy consumption model for each elastic task.
Finally, Eq. \eqref{eq:mec-const} enforces the maximum computation speed on the MEC unit.
Note that $z(f_k^{mec})$ denotes the computational speed allocated to user $k$ on the MEC unit. 

\subsection{Challenges, Objective Trade-offs, and Pareto Frontier}
The optimization problem in Eq. \eqref{eq:ee-maximization} aims to maximize a weighted sum of video quality, task response time, and energy consumption while limiting the task response time to a certain deadline.
There are several key challenges to solve the optimization problem in Eq. \eqref{eq:ee-maximization}, including:
\emph{(i) Multi-User Shared Environment:} There are several users in the environment with different communication, computation, and deadline requirements.
As denoted in Eq. \eqref{eq:mec-const}, VR users share the MEC's computational resources among them.
This means that a change in one user's communication and computation requirements may lead to a performance degradation for other users. This is due to the fact that the MEC unit has limited computational resources that need to be allocated to a subset of users to improve the overall system performance.
\emph{(ii) Time-Varying System Characteristics:} 
The environment is dynamic due to varying $360^\circ$ video characteristics and time-varying network conditions.
The amount of computational resources required to compute each elastic task varies from one task to another depending on user's request.
Moreover, each communication channel has different characteristics (\eg 5G wireless technology enables multi-Gbps transmission rates at the expense of highly dynamic channels due to blockage and mobility, while 4G wireless technology provides a less dynamic channel at the expense of a lower transmission rate).
\emph{(iii) Partially Observable Environment:} 
The optimization problem in Eq. \ref{eq:ee-maximization} needs to be solved every time a new task arrives at the headset, while some task features (\eg transmission energy consumption, transmission time, etc.) are not observable in the decision-making stage.

\textbf{Optimal Solution:}\label{sec:opt-sol}
To gain a deeper insight into the structure of the optimal solution set of the multi-user, multi-objective optimization problem described in Eq. \ref{eq:ee-maximization}, we first constructed the objective space through an exhaustive brute-force approach.
Specifically, we uniformly sampled one million combinations of the objective weights $(w_0, w_1, w_2)$, where $0 \le w_i \le 1$ for all $i \in \{0, 1, 2\}$. For each sampled weight vector, the optimization problem was solved to obtain the corresponding average video quality, response time, and energy consumption.
This brute-force exploration ensures that the full range of trade-offs among the three competing objectives is captured, independent of any learning-based approximation or algorithmic biases.

\begin{figure*}[t]
    \centering
    \begin{subfigure}{0.32\linewidth}
        \centering
        \includegraphics[width=.99\columnwidth]{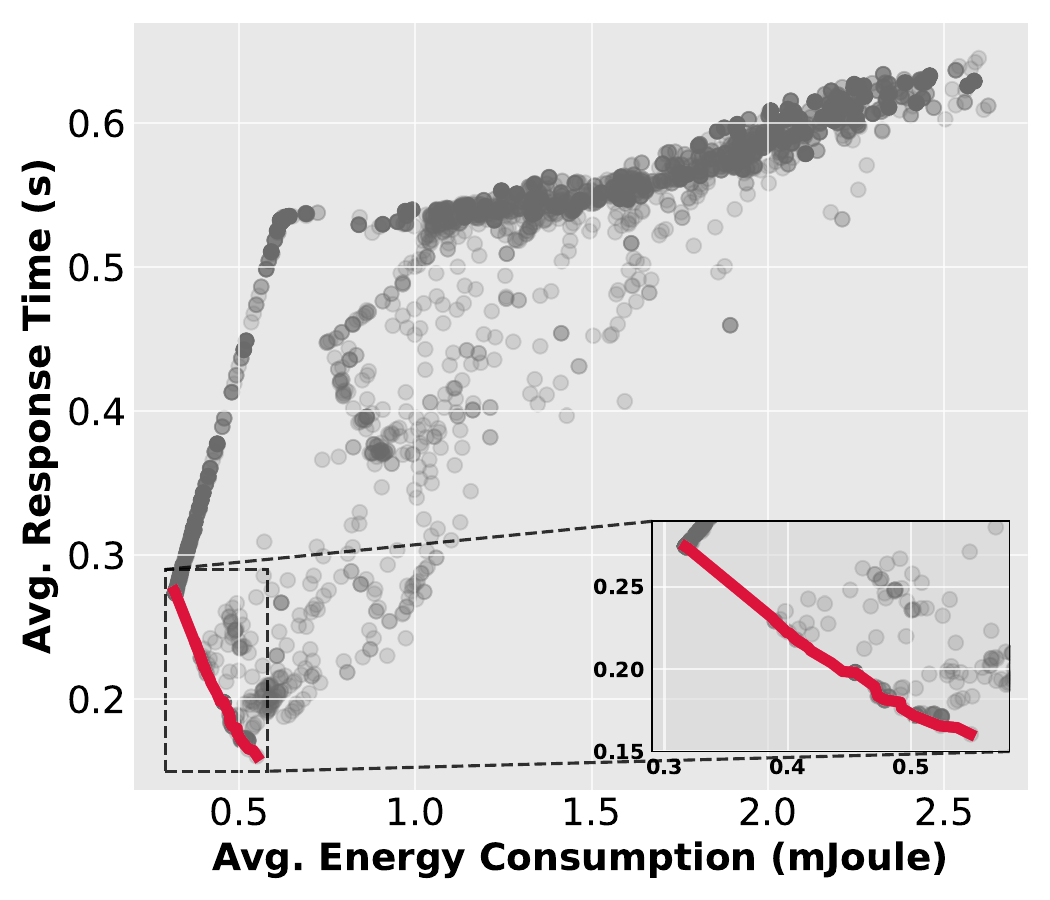}
        \subcaption{}
    \end{subfigure}
    \begin{subfigure}{0.32\linewidth}
    \centering
        \includegraphics[width=.99\columnwidth]{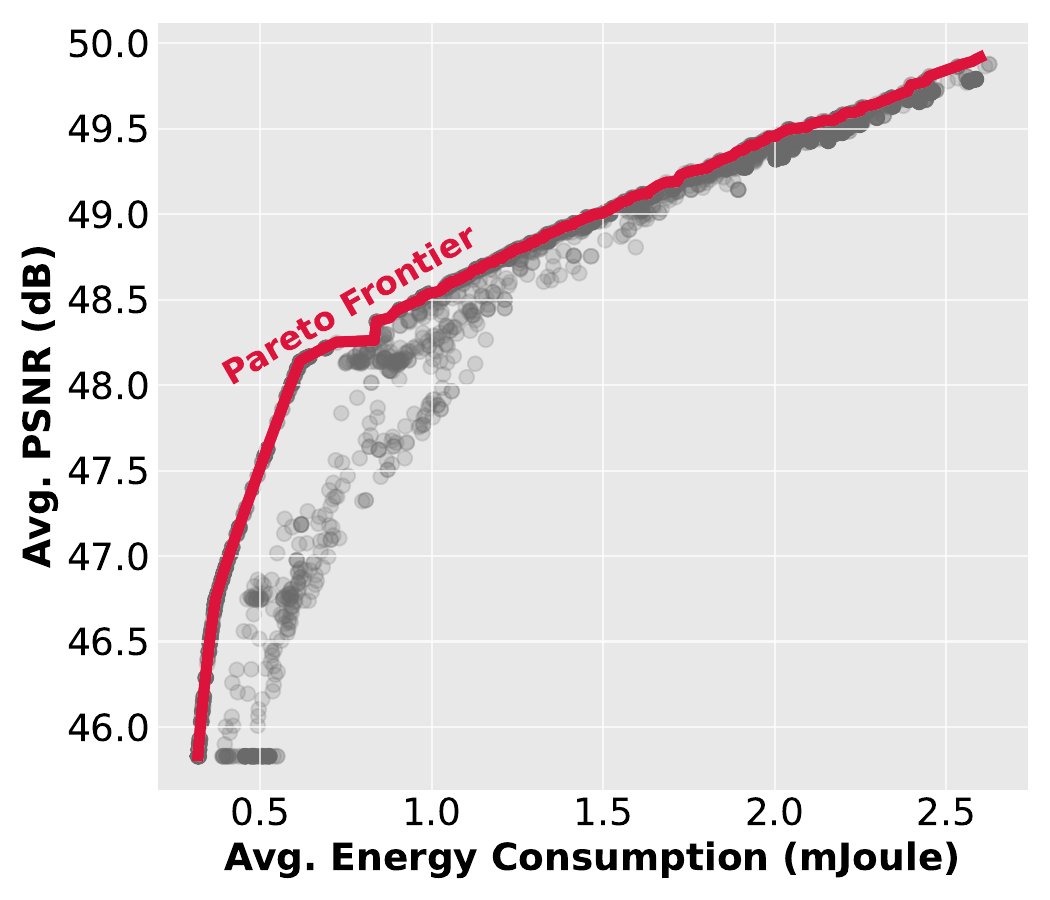}
        \subcaption{}
    \end{subfigure}
    \begin{subfigure}{0.32\linewidth}
    \centering
        \includegraphics[width=.99\columnwidth]{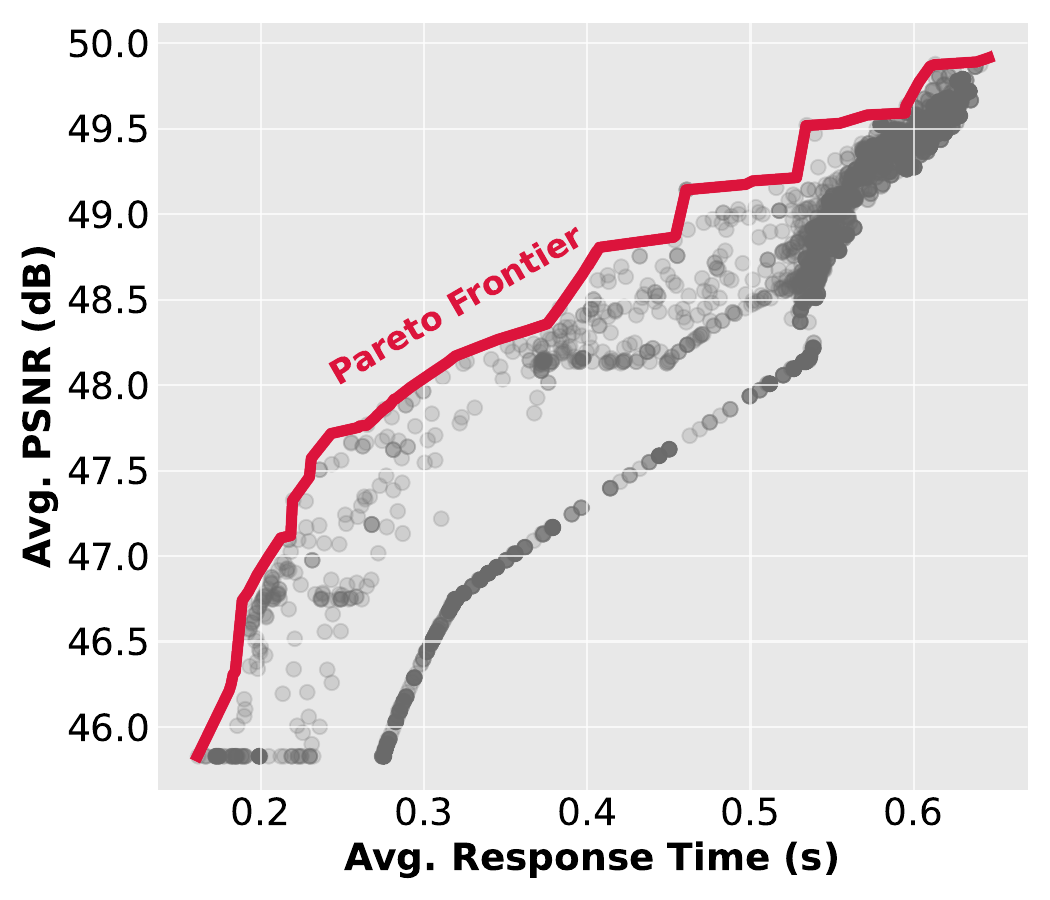}
        \subcaption{}
    \end{subfigure}
    \caption{
    Objective space of the optimization problem with $0 \le w_i \le 1, \forall i \in \{0, 1, 2\}$ over 1,000,000 set of weights. 
    Each point corresponds to a feasible solution obtained for a sampled weight vector, and the Pareto-optimal frontiers are highlighted as the boundary of solutions that cannot be improved in one objective without sacrificing at least one of the others.
    (a) Trade-off between response time and energy consumption, (b) Trade-off between PSNR and energy consumption, (c) Trade-off between PSNR and response time.
    }
    \label{fig:obj-space}
\end{figure*}

The resulting objective space is depicted in Fig. \ref{fig:obj-space}. Each point in the figure corresponds to a feasible solution obtained for a particular weight combination, and the Pareto frontiers are highlighted as the boundary of solutions that cannot be further improved in one objective without sacrificing the other.
The results clearly reveal the inherent trade-offs in the system: lower energy consumption often comes at the expense of increased response time, while higher PSNR values require additional computation and transmission resources, thereby increasing both response time and energy consumption.
The Pareto frontiers shown in Fig. \ref{fig:obj-space} serve as a benchmark to evaluate the effectiveness of our proposed algorithms, since any near-optimal policy should achieve a performance close to these boundaries.

\section{Constrained Multi-Agent Decision-Making Framework}\label{sec:solution}
\bb{
Our goal is to develop a dynamic decision-making algorithm that addresses the challenges mentioned above.
Learning based decision-making methods, DRL in particular, have shown promising results in solving decision-making tasks in various applications.
The success of the learning-based methods stems from the fact that they do not rely on predefined models of the environment, but instead learn the optimal policy through repeated interactions with the environment and collecting rewards that are commensurate with the quality of the actions.
}
To tackle the above-mentioned challenges, we develop a MARL agent. 
The objective of each agent in a multi-agent setting is to maximize its expected cumulative reward:
\begin{equation}
\label{eq:return-func}
    J(\pi) = \mathbb{E} 
    \Bigg[
        \sum_{t=0}^{\infty} \gamma^t r(\mathbf{s_t}, \mathbf{a_t}) | \pi = (\pi_k, \pi_{-k})
    \Bigg].
\end{equation}
Here, $\pi$ is the joint policy (\ie policy of all agents in the environment), $\pi_k$ is the agent $k$'s policy, $\pi_{-k}$ represents the policies of all other agents, $r(\mathbf{s_t}, \mathbf{a_t})$ is the reward signal received by \bb{all agents} at time $t$ based on the current global state $s_t$ and joint action $\mathbf{a}_t = [a^1_t, a^2_t, ..., a^K_t]$, and $\gamma \in [0, 1)$ is the discount factor for future rewards. 

In the rest of this section, 
we present two multi-agent reinforcement learning architectures for elastic task offloading: the Centralized Phasic Policy Gradient (CPPG)~\cite{amato2024intro} and the Independent Phasic Policy Gradient (IPPG).
These two methods embody distinct design philosophies that reflect the broader paradigms of multi-agent reinforcement learning.
Specifically, CPPG adopts a centralized training and centralized execution (CTCE) approach, where a single agent at the MEC unit has full observability of all users’ states and makes a joint offloading decision for the entire system. 
This design enables CPPG to directly capture the coupling between users’ communication and computation demands, leading to globally coordinated decisions. However, the requirement of full system observability and centralized decision-making increases the computational burden at the MEC and may limit scalability in highly dynamic settings.

\begin{figure}[t]
    \centering
    \includegraphics[width=0.7\linewidth, trim= 6mm 0mm 0mm 0mm, clip=true]{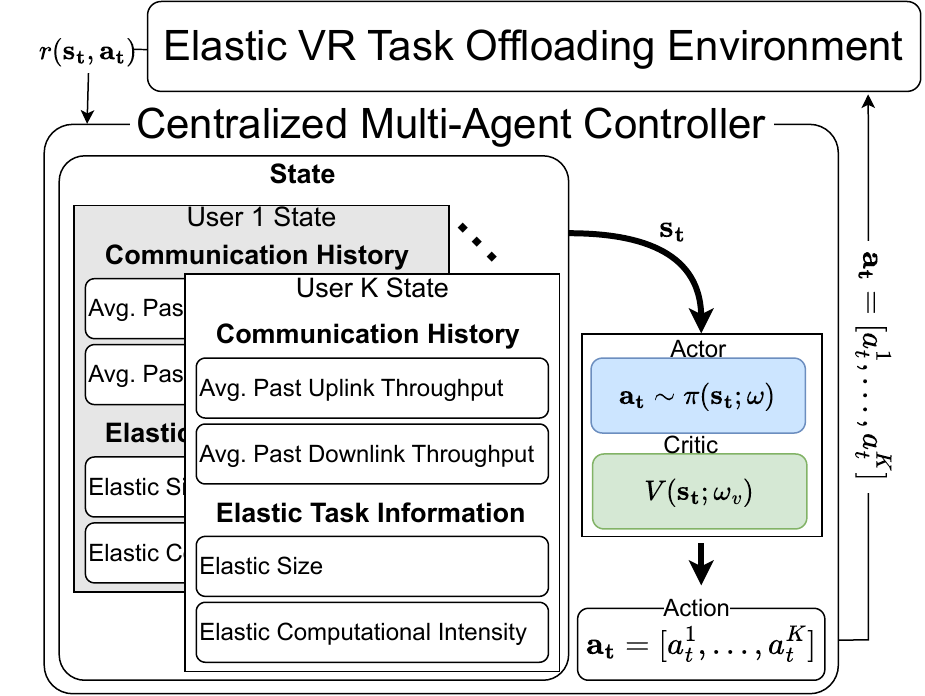}
    \caption{Illustration of the CPPG framework. A centralized controller leverages the global state information of all users (including uplink/downlink history, task size, computational intensity, and communication history) to coordinate multi-agent elastic task offloading and resource allocation decisions.}
    \label{fig:ctce}
\end{figure}

In contrast, IPPG follows a centralized training with decentralized execution (CTDE) paradigm. 
During training, agents leverage shared information and parameter sharing to learn robust policies, but during execution, each user applies its own local policy based solely on its local state.
This decentralized execution reduces the communication and computation overhead of centralized decision-making and improves scalability.
At the same time, the lack of full observability during execution makes coordination among users more challenging, which may lead to sub-optimal performance compared to CPPG.
Together, CPPG and IPPG highlight the fundamental trade-off between coordination accuracy and scalability in multi-agent task offloading: CPPG achieves stronger global coordination at the cost of centralized complexity, while IPPG offers more practical deployment through independent execution with potentially reduced performance.
A schematic comparison of the two architectures is provided in Fig.~\ref{fig:ctce} and Fig.~\ref{fig:ctde}, which illustrate the difference between centralized versus decentralized execution.

\subsection{Centralized Phasic Policy Gradient (CPPG)}\label{sec:cppg}
In this section, we present a centralized training and execution approach, called CPPG, for solving Eq. \eqref{eq:ee-lagrange}.
First, we describe the decision-making flow for each task and then present the details of the CPPG agent (\ie state, action, reward, and architecture of the neural network). Finally, we delve into the details of the training process.

\textbf{Decision-Making Flow and State:}
At each time step, the CPPG agent observes the state of the environment $\mathbf{s_t} = [s_t^1, s_t^2, .., s_t^K]$, which comprises the state of each agent in the environment.
The $k^{th}$ agent's state $s_t^k$ includes \emph{task information and average past network throughputs}.
The elastic task information includes elastic task size $\big[S_k(0), S_k(1), ..., S_k(L)\big]$ and task computational intensity $\big[I_k(0), I_k(1), ..., I_k(L)\big]$.

\textbf{Action:}
Once the state $\mathbf{s_t}$ is observed, the CPPG agent takes a joint multi-task action $\mathbf{a_t} = [a_t^1, a_t^2, .., a^K_t]$, where
$a^k_t = (e_k^t, u_k^t)$ is the elasticity parameter and offloading decision, as described in Eq. \eqref{eq:action}, of the $k^{th}$ user at time $t$.


\textbf{Reward:}
After executing the joint action, the environment transitions to a new state and each agent receives a reward $r^k_{t+1}$.
The goal of our agent is to maximize the objective function in Eq. \eqref{eq:ee-maximization}, \bb{while satisfying the deadline constraints.
MARL methods are generally capable of capturing the uncertainty in the system due to varying task computational and communication requirements and time-varying network conditions.
However, MARL methods, in general, are not well-suited for optimization problems with inequality constraints.
Thus, we need to incorporate Eq. \eqref{eq:deadline-const} into the objective function of Eq. \eqref{eq:ee-maximization}.
Leveraging the duality principle, 
we modify the optimization problem to capture the constraint violation and penalize the MARL agent for not meeting the constraints, which leads to the Lagrangian dual problem:
\begin{equation}
\label{eq:ee-lagrange}
\begin{aligned}
    \min_{\{\lambda_k\}} \max_{\mathbf{e}, \mathbf{u}} &  \quad
            \mathit{QTE}(\mathbf{e}, \mathbf{u})
            + \sum_{k=1}^{K} \lambda_k \Big(T^d_k - T^r_k(e_k, u_k)\Big)
             \\
    \text{s.t.} & \quad \text{Eqs.} \;\; \eqref{eq:tr-const} \text{, } \eqref{eq:energy-const} \text{, and } \eqref{eq:mec-const}
\end{aligned}
\end{equation}
which is equal to:
\begin{equation}
\label{eq:qoe-lagrange}
\begin{aligned}
    \min_{\{\lambda_k\}} &  \quad
            \mathit{QTE}(\mathbf{e}^{\ast}, \mathbf{u}^{\ast})
            + \sum_{k=1}^{K} \lambda_k \Big(T^d_k - T^r_k(e_k^{\ast}, u_k^{\ast})\Big)
             \\
    \text{s.t.} & \quad \text{Eqs.} \;\; \eqref{eq:tr-const} \text{, } \eqref{eq:energy-const} \text{, and } \eqref{eq:mec-const}
\end{aligned}
\end{equation}
Here, $\mathbf{u}^{\ast}$ and $\mathbf{e}^{\ast}$ are the optimal offloading decision and elasticity parameter, respectively.
$\lambda_k$ is a penalty coefficient for not meeting the deadline of the elastic task in hand, which is obtained by:
\begin{equation}
\label{eq:lambda-coeff}
\begin{aligned}
    \lambda_k^{\ast} = \argmin_{\lambda_k} \;\; \lambda_k \Big(T^d_k - T^r_k(e_k^{\ast}, u_k^{\ast})\Big).
\end{aligned}
\end{equation}
This means that the $\lambda_k^{\ast}$ can easily be achieved by evaluating $T^d_k - T^r_k(e_k^{\ast}, u_k^{\ast})$:
\begin{equation}\label{eq:lambda-coeff-res}
    \lambda_k^{\ast} = 
    \begin{cases}
        0 \quad
        &T^d_k - T^r_k(e_k^{\ast}, u_k^{\ast}) \ge 0 \\
        \lambda_k^{penalty} \Big(T^d_k - T^r_k(e_k^{\ast}, u_k^{\ast})\Big) \quad 
        &T^d_k - T^r_k(e_k^{\ast}, u_k^{\ast}) < 0.
    \end{cases}
\end{equation}
Therefore, by employing Eq. \eqref{eq:lambda-coeff-res} to obtain $\lambda_k^{\ast}$, the MARL agent is penalized for not meeting the task deadline. Note that the penalty is proportional to the amount of deadline violation.

The objective function in Eq. \eqref{eq:ee-lagrange} accounts for deadline violation constraints.
Moreover, it allows us to set $\gamma = 0$ in Eq. \eqref{eq:return-func} because (i) tasks are independent across different steps, (ii) MEC CPU allocation is re-solved per step with no backlog carry-over, and (iii) rewards are per-task and memoryless.
}
Thus, all of the agents must cooperate/compete \bb{(\ie mixed game)} to maximize the discounted sum of the following reward function:
\begin{equation} \label{eq:reward-func}
    r (\mathbf{s_t}, \mathbf{a_t}) =  
        \mathit{QTE}(\mathbf{e}, \mathbf{u})
            + \sum_{k=1}^{K} \lambda_k \big[T^d_k - T^r_k(e_k^t, u_k^t)\big]\indicator_{\{T^d_k < T^r_k(e_k^t, u_k^t)\}}.
\end{equation}
\bb{
This reward captures the performance gain or loss as a result of the action (\ie elasticity parameter and offloading decision) by the MARL agent, thereby enabling the agent to learn the actions that lead to improvement in the average weighted sum of video quality, task response time, and energy-efficiency.
}


\textbf{CPPG Architecture:}
The CPPG agent, as shown in Fig. \ref{fig:ctce}, is composed of an actor network $\omega$ and a critic network $\omega_v$.
The actor network outputs policies of all agents along with a set of auxiliary values that estimate the state value.
The critic network outputs the estimated state value.
This architecture reduces the interference between policy and value loss, while distilling features from the value function into the policy network \cite{cobbe-2020-ppg}.
Both actor and critic networks are composed of a convolutional layer and a dense layer. 
We employ a two-phase training procedure, consisting of a policy training phase, and an auxiliary training phase~\cite{cobbe-2020-ppg, Deheng-2020-Mastering, Tianchi-2023-Buffer}.


\textbf{Policy Optimization:}
In the policy training phase, we update the actor and critic networks.
The policy network is trained by dual-clip proximal policy optimization (PPO) \cite{schulman-2017-proximal}:
\begin{equation}
\label{eq:d-clip-loss}
    \begin{aligned}
        \mathcal{L}^{DClip} = \hat{\mathbb{E}}
            \Big[ \indicator(\hat{A}_t < 0)\max(\mathcal{L}^{PPO}, c\hat{A}_t) + 
            \indicator(\hat{A}_t \geq 0)\mathcal{L}^{PPO} \Big].
    \end{aligned}
\end{equation}
Here, $\indicator(.)$ is a binary indicator function and $\hat{A}_t = r_t + \gamma V_{\omega_v}(\mathbf{s_{t+1}}) - V_{\omega_v}(\mathbf{s_t})$ is the advantage function calculated based on the current state value estimate and the discount factor $\gamma = 0$, and $\mathcal{L}^{PPO}$ is the surrogate vanilla PPO loss:
\begin{equation}
\label{eq:ppo-loss}
    \begin{aligned}
        \mathcal{L}^{PPO} = \mathcal{L}^{Clip}(\pi_{\omega}, \hat{A}_t) + \beta H_{\omega}(\mathbf{s_t}) + \mathcal{L}^{Value},
    \end{aligned}
\end{equation}
where $H_{\omega}(\mathbf{s_t})$ is the entropy of all the policies and $\beta$ is the entropy weight.
The entropy loss and its associated weight balance the trade-off between exploration and exploitation in the learning process.
$\mathcal{L}^{Value}$ is the critic network loss:
\begin{equation}
\label{eq:value-loss}
    \begin{aligned}
        \mathcal{L}^{Value} &= \mathbb{E}\left[ 
            \frac{1}{2}
            (V_{\omega_v}(\mathbf{s_t}) - V_{targ}(\mathbf{s_t}))^2 
        \right],
    \end{aligned}
\end{equation}
and $\mathcal{L}^{Clip}$ is the single-clip policy loss defined as: 
\begin{equation}
\label{eq:clip-loss}
    \begin{aligned}
        \mathcal{L}^{Clip} = \min \Big[ &\rho(\omega, \omega_{old}) \hat{A}_t,  
  clip(\rho(\omega, \omega_{old}), 1 \pm \varepsilon) \hat{A}_t \Big].
    \end{aligned}
\end{equation}
Here $\rho(\omega, \omega_{old}) \hat{A}_t$ measures the gained/lost reward as the joint policy changes, which determines how a change in one agent's policy affects other agents reward, (\ie video quality, response time, and energy-consumption).
The change in the joint policy w.r.t. the old policy is measured by multiplying the changes in each agent's policy.
\begin{equation}
\label{eq:rho-i}
    \begin{aligned}
        \rho(\omega, \omega_{old}) 
        = \frac{
                \pi (\mathbf{a_t}|\mathbf{s_t}; \omega)
            }{
                \pi (\mathbf{a_t}|\mathbf{s_t}; \omega_{old})
            }
        = \prod^{K}_{k=1}
            \frac{
                \pi_k (a_t^k|\mathbf{s_t}; \omega)
            }{
                \pi_k (a_t^k|\mathbf{s_t}; \omega_{old})
            }.
    \end{aligned}
\end{equation}

\textbf{Auxiliary Training Phase:}
In the auxiliary phase, we further optimize the actor and critic networks according to a joint loss function $\mathcal{L}^{Joint}$ on all the experiences.
The joint loss function $\mathcal{L}^{Joint}$ is composed of a behavioral cloning loss and an auxiliary value loss:
\begin{equation}
\label{eq:joint-loss}
    \begin{aligned}
        \mathcal{L}^{Jonit} = 
        \mathbb{E}
        \Bigg[ 
            \sum_{k=1}^K 
            \mathbf{KL} \left(\pi_k (\mathbf{s_t}; \omega_{old}), \pi_k (\mathbf{s_t}; \omega)\right)
        \Bigg] 
        + \mathcal{L}^{Aux}.
    \end{aligned}
\end{equation}
Here, $\mathbf{KL}(.,.)$ is the behavioral cloning loss, representing the KL-divergence between the original policy and the updated policy.
The $\mathcal{L}^{Aux}$ in Eq. \eqref{eq:joint-loss} updates the auxiliary value by minimizing the mean squared loss function:
\begin{equation}
\label{eq:aux-loss}
    \begin{aligned}
        \mathcal{L}^{Aux} &= \mathbb{E}\left[ 
            \frac{1}{2}
            (V_{\omega}(\mathbf{s_t}) - V_{targ}(\mathbf{s_t}))^2 
        \right].
    \end{aligned}
\end{equation}


\textbf{Training Algorithm:}
Algorithm~\ref{alg:cppg-alg} presents the training process of the CPPG agent.
The training continues for multiple iterations until convergence.
Each iteration consists of three phases.
In the first phase, we perform the current policy $\pi_{\omega}$ on a randomized environment to collect new experiences (lines 3-7).
We first sample the joint action, and then apply the chosen action on the edge-assisted VR elastic task offloading environment.
Then, we store the resulting transition in a replay buffer for training purposes.
In the second phase (\ie policy training phase), we update both actor and critic networks.
We use a random batch from replay buffer to compute the dual-clip PPO loss $\mathcal{L}^{DClip}$, and update the networks (lines 18-20).
Finally, in the third phase (\ie auxiliary training phase), we further update the actor and critic networks by optimizing the behavioral cloning and value losses using all the replay buffer data (lines 21-24).

\begin{algorithm}[t]
\caption{CPPG \& IPPG Training Process}
\label{alg:cppg-alg}
\begin{algorithmic}[1]
    \Require $mode \in \{\text{Centralized},\; \text{DeCentralized} \}$
    \State Initialize $\mathcal{B} \gets \varnothing$
    \For{task $=1,2,\dots$}
        \If{$mode == \text{Centralized}$}
            \State $\mathbf{a_t} \sim \pi (\mathbf{a_t}|\mathbf{s_t}; \omega)$ \Comment{Elasticity parameter and offloading decisions are determined.}
            \State $\mathbf{s_{t+1}}, r_{t+1} \gets \text{ACT}(\mathbf{s_t}, \mathbf{a_t})$ \Comment{The joint action is applied to the environment.}
            \State $\mathcal{B} \gets \mathcal{B} \cup \{\mathbf{s_t}, \mathbf{a_t}, r_{t+1}, \mathbf{s_{t+1}}, V_{targ}(\mathbf{s_t})\}$  \Comment{The transition is recorded for training.}
        \Else
            \For{$k=1...K$}
                \State $a^k_t = (e_k^t, u_k^t) \sim \pi (a_t^k|s_t^k; \omega)$ \Comment{Elasticity parameter and offloading decisions are determined for user $k$.}
            \EndFor
            \State $\mathbf{a_t} = [a_t^1, a_t^2, .., a^K_t]$
            \State $\mathbf{s_{t+1}}, r_{t+1} \gets \text{ACT}(\mathbf{s_t}, \mathbf{a_t})$ \Comment{The joint action is applied to the environment.}
            \For{$k=1...K$}
                \State $\mathcal{B} \gets \mathcal{B} \cup \{s^k_t, \mathbf{a_t}, r_{t+1}, s^k_{t+1}, V_{targ}(s^k_t)\}$ \Comment{The transitions are recorded for training.}
            \EndFor
        \EndIf
        \If{task $\% \; N_{update} = 0$}
            \For{$i = 1,2,\dots, N_{Policy}$}  \Comment{Policy training phase}
                \State Optimize $\mathcal{L}^{DClip}$ w.r.t. $\omega, \omega_{v}$ (\textit{i.e.}, Eq. \eqref{eq:d-clip-loss})
            \EndFor
            \For{$i = 1,2,\dots, N_{aux}$} \Comment{Auxiliary training phase}
                \State Optimize $\mathcal{L}^{Value}$ w.r.t. $\omega_v$  (\textit{i.e.}, Eq. \eqref{eq:value-loss})
                \State Optimize $\mathcal{L}^{Joint}$ w.r.t. $\omega$  (\textit{i.e.}, Eq. \eqref{eq:joint-loss})
            \EndFor
            \State $\mathcal{B} \gets \varnothing$ \Comment{Clear replay buffer}
            \State $\omega_{old} \gets \omega$, $\omega_{v_{old}} \gets \omega_v$ \Comment{Update target network}
        \EndIf
    \EndFor
\end{algorithmic}
\end{algorithm}

    


\subsection{Independent Phasic Policy Gradient}\label{sec:IPPG}
\begin{figure}[t]
    \centering
    \includegraphics[width=0.7\linewidth, trim= 6mm 0mm 25mm 0mm, clip=true]{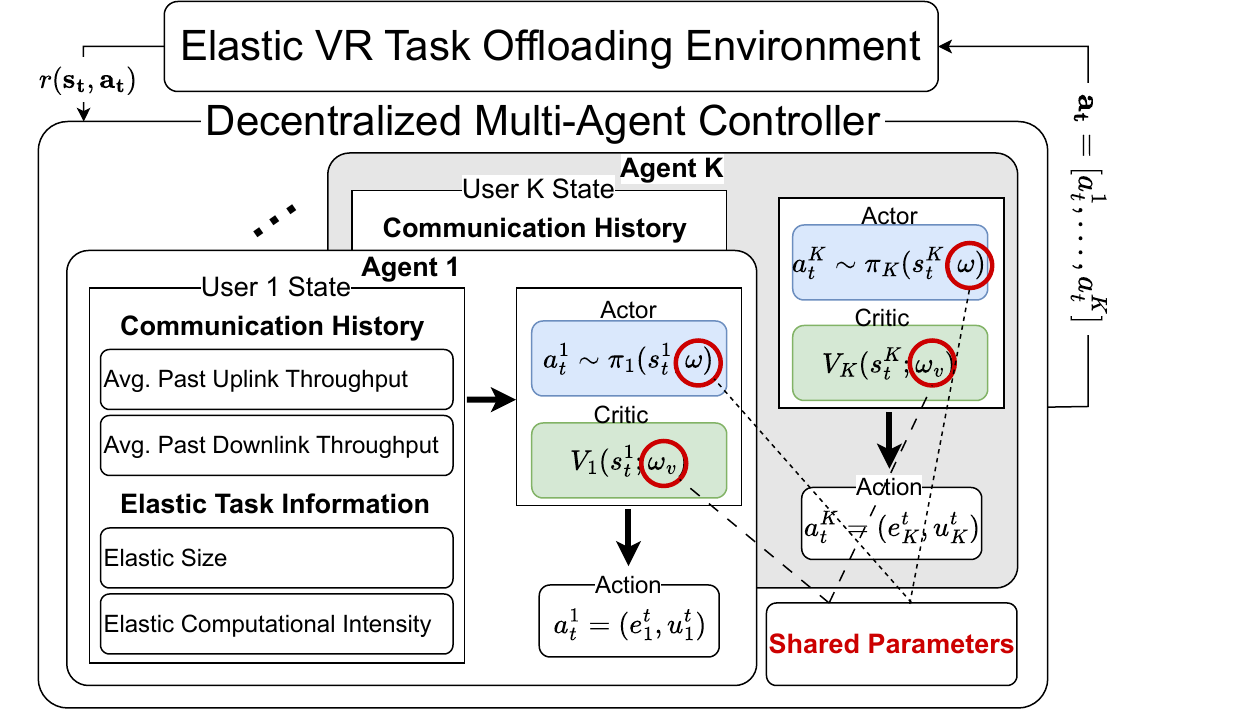}
    \caption{Illustration of the IPPG framework. Each agent (Actor–Critic pair) makes task offloading decisions based solely on its own local state (uplink/downlink history, task size, computational intensity, and communication history) without centralized coordination, forming a decentralized multi-agent controller}
    \label{fig:ctde}
\end{figure}

To extend the CPPG framework to a decentralized setting, we present IPPG method to solve Eq. \eqref{eq:ee-lagrange}, as depicted in Fig.~\ref{fig:ctde}.
Similar to CPPG, IPPG is composed of an actor network $\omega$ and a critic network $\omega_v$. 
However, unlike CPPG, both actor and critic only observe the state of one user at a time, which means $\mathbf{s_t} = s_t^k$ (compared with $\mathbf{s_t} = [s_t^1, s_t^2, .., s_t^K]$ defined in Sec. \ref{sec:cppg}).
Unlike CPPG, the state space, in this architecture, does not depend on the number of users in the system, which in turn provides multiple benefits such as i) higher sample efficiency, ii) decentralized decision-making~\cite{Yu-2022-PPO}, and iii) faster training.

\textbf{Policy Optimization:}
Similar to CPPG, we employ a two-phase training procedure for IPPG agent. 
However, the change in the architecture leads to a few changes in the loss functions.
Eq. \ref{eq:rho-i}, which measures the change in the joint policy, is not a function of joint policy anymore since the actor only decides on the action of one agent at a time. Subsequently, the change in the policy is measured via:
\begin{equation}
    \begin{aligned}
        \rho(\omega, \omega_{old}) 
        = 
            \frac{
                \pi (a_t^k|s_t^k; \omega)
            }{
                \pi (a_t^k|s_t^k; \omega_{old})
            }.
    \end{aligned}
\end{equation}

In the auxiliary training phase, the behavioral cloning loss, as described in Eq. \eqref{eq:joint-loss}, and the auxiliary value loss, as described in Eq. \eqref{eq:aux-loss} do not change. Note that this is due to the way we have defined the state space in this setting.

\begin{figure*}[ht]
    \centering
    \begin{subfigure}{0.32\linewidth}
        \centering
        \includegraphics[width=.99\columnwidth]{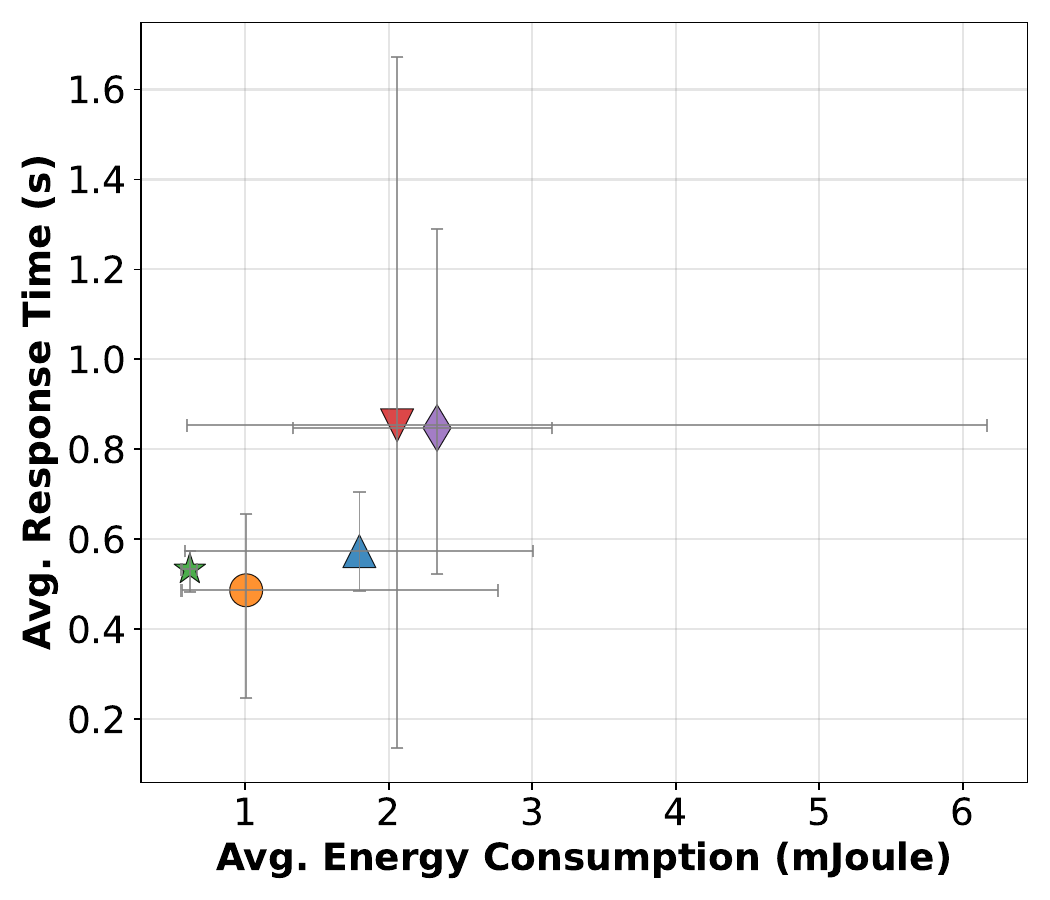}
        \subcaption{}
    \end{subfigure}
    \begin{subfigure}{0.32\linewidth}
    \centering
        \includegraphics[width=.99\columnwidth]{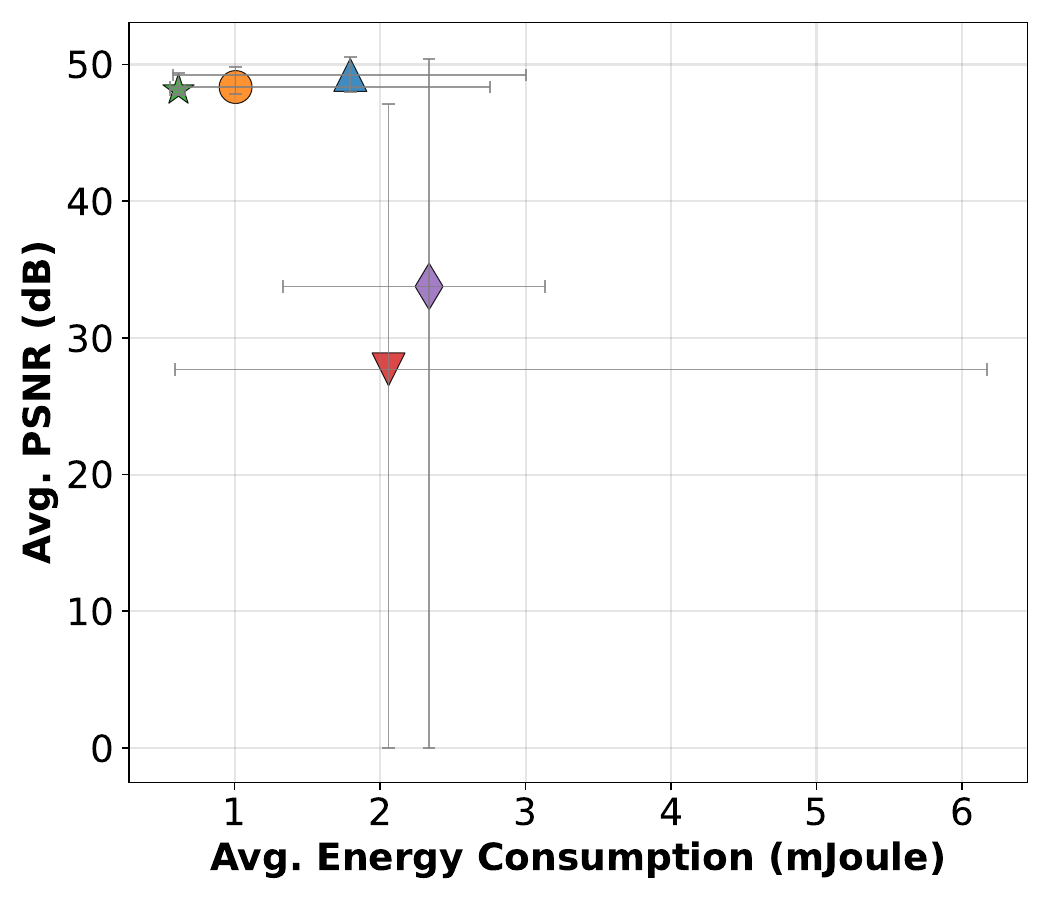}
        \subcaption{}
    \end{subfigure}
    \begin{subfigure}{0.32\linewidth}
    \centering
        \includegraphics[width=.99\columnwidth]{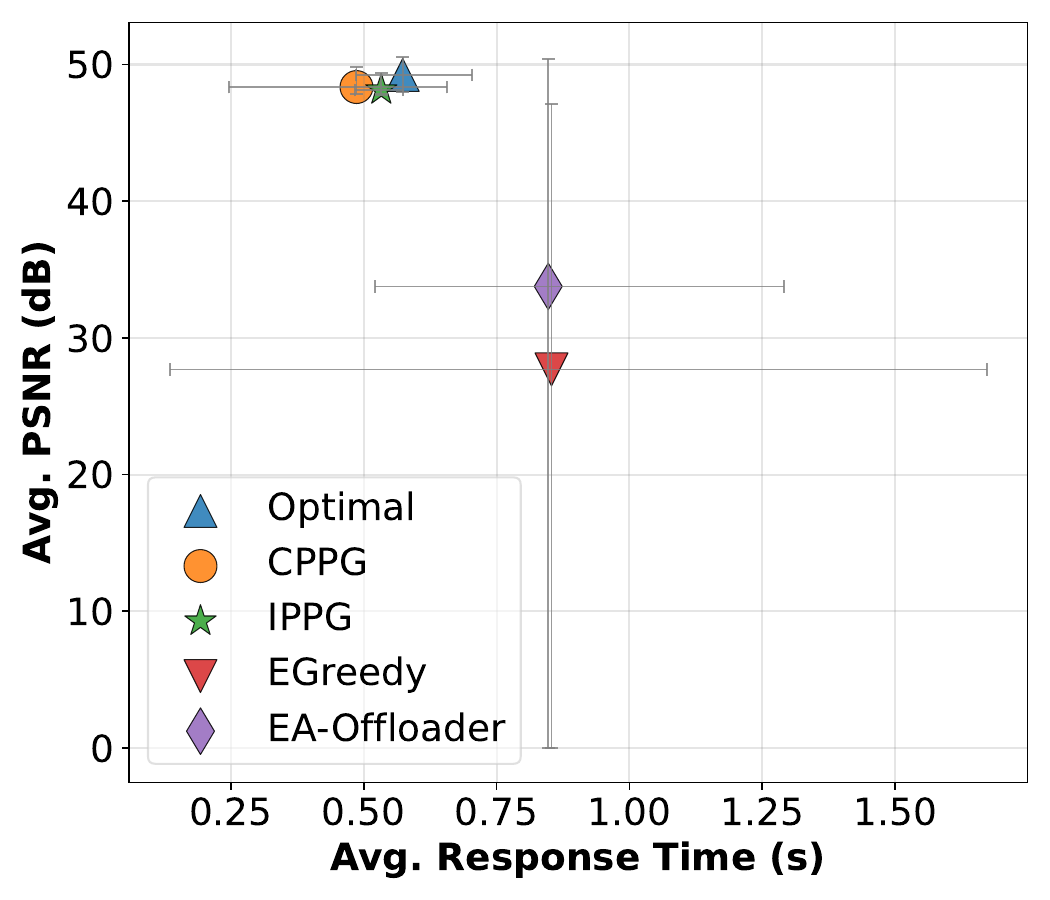}
        \subcaption{}
    \end{subfigure}
    \caption{
    Performance trade-offs between PSNR, latency, and energy consumption in the testing stage for an environment with three users compared to the optimal achievable performance. $w_0=0.35$, $w_1=0.85$,
    $w_2=0.15$. 
    }
    \label{fig:deployment-performance}
\end{figure*}


\section{Evaluation}\label{sec:evaluation}
We evaluate our proposed framework through extensive simulations against three baselines: the optimal solution developed in Sec. \ref{sec:opt-sol}, a Neural $\epsilon$-Greedy, refereed to as EGreedy algorithm, and an elasticity-agnostic learning-based offloading method, dubbed EA-Offloader.
The Neural $\epsilon$-Greedy algorithm represents a class of task offloading approaches that leverage multi-armed bandit methods to learn offloading policies \cite{Yang-2022-Multi, Zhu-2019-BLOT, Wang-2022-Decentralized}.
We extend this baseline to jointly make offloading and elasticity decisions, while also employing parameter sharing during training to improve learning efficiency.
The EA-Offloader method adopts Proximal Policy Optimization (PPO) \cite{schulman-2017-proximal} to learn task offloading decisions. 
In the EA-Offloader baseline, the elasticity parameter is fixed to $e_k=4$ for all users, meaning the method only learns the offloading policy without adapting task elasticity.
This approach, adapted from prior task offloading literature \cite{Tang-2022-Deep, Zhou-2022-Online, Cozzolino-2023-Nimbus}, therefore does not capture the elasticity dimension of the problem.

\begin{figure*}[t]
    \centering
    \includegraphics[width=.6\columnwidth]{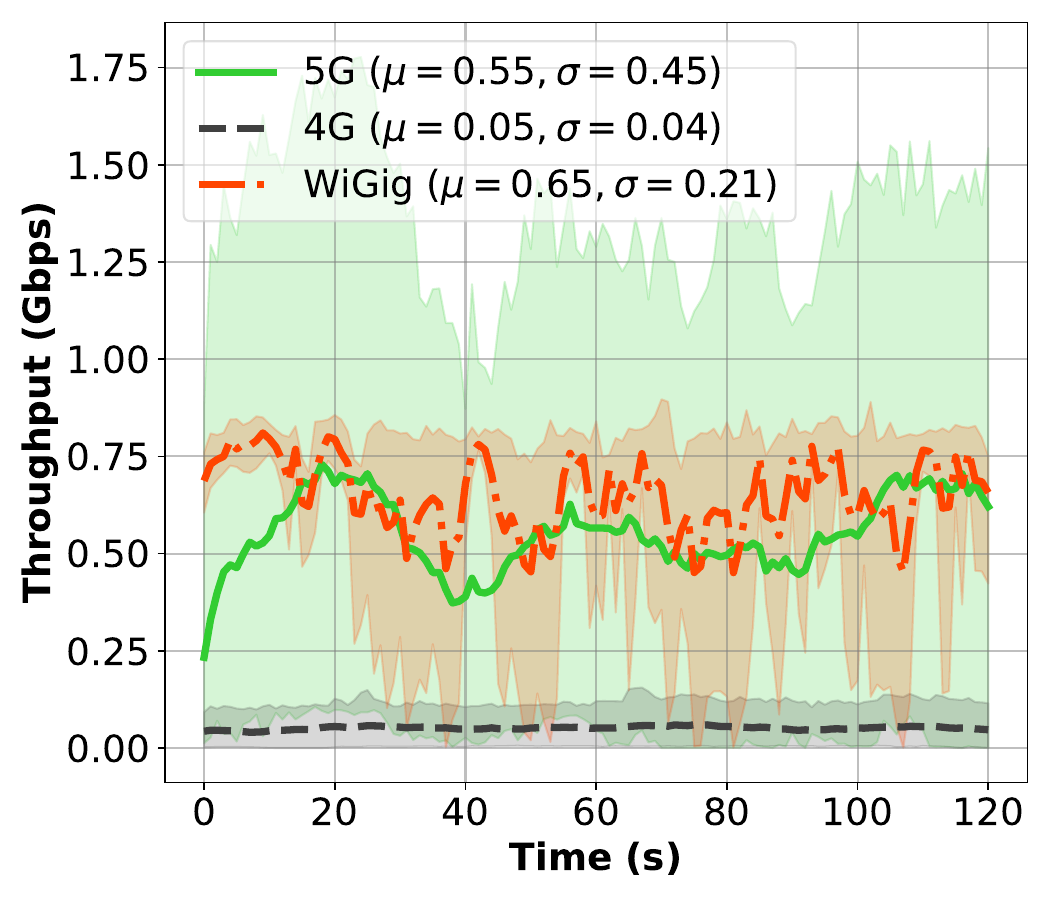}
    \caption{
    Statistics of 4G, 5G, and WiGig network traces used in our simulation.
    }
    \Description[Short description for accessibility]{Longer, detailed description of the image content.}
    \label{fig:eda-c}
\end{figure*}


In our simulation, we consider a multi-connectivity VR system, where each user has access to three communication channels operating on 4G, 5G, and WiGig technologies.
VR users stream full UHD $360^\circ$ videos taken from a publicly available $360^\circ$ video dataset \cite{Chakareski-2021-Full}.
This dataset includes 15 videos with various spatio-temporal characteristics.
Each video is divided into 36 segments of fixed time duration.
Each video segment is a sequence of video frames, and seven \bb{($L=7$)} layers of increased immersion fidelity for each frame is available.
The elastic task that we consider is to decode each video segment.
Moreover, we use a dataset of 4G, 5G, and WiGig network throughput traces.
The 4G and 5G traces were collected in two different cities in the U.S. and from commercial operators (T-Mobile and Verizon)~\cite{Narayanan-2021-Variegated}.
Fig. \ref{fig:eda-c} reports the statistical characteristics of these network traces over time.
\bb{Transmission power settings for our simulations follow measurement studies presented in \cite{Saha-2017-detailed, Narayanan-2021-Variegated} as reported in Table \ref{tab:parameter-values}.}
We employ these datasets to train the IPPG agent for $2,000$ episodes.
However, the CPPG agent requires more samples for convergence due to its larger state space. Thus, to ensure that CPPG has access to the same number of samples, we scale the training episodes with the number of user (\eg $10,000$ training episodes for an environment with $K=5$).
To ensure a fair comparison, we use the same configuration for all the baselines as presented in Table \ref{tab:parameter-values}.
\begin{table}[t]
    \centering
    \caption{Simulation and training parameter values.}
    \label{tab:parameter-values}
    \resizebox{.5\linewidth}{!}{
    \begin{tabular}{lc}
        \toprule
        Definition/Explanation & Parameter \& Value \\
        \midrule
        Number of VR users & $K \in \{2, 3, 4, 5, 6, 7, 8\}$ \\
        Tasks deadline & $T_d^k = 1$ Second\\
        \midrule
        CPU capacitance factor & $\kappa = 1\mathrm{e}{-27}$ \\
        Users' CPU frequency & $f_k^{vr} = 2.4 \,GHz$ \\
        User computing speed & $Z_{k}=200 \, Mbps$ \\
        MEC computing speed & $Z_{mec} = 12 \,Gbps$ \\
        \midrule
        5G transmission power & $5.27 \, mW/Mbps$ \\
        4G transmission power & $57.99 \, mW/Mbps$ \\
        WiGig transmission power & $6.15 \, mW/Mbps$ \\
        \midrule
        Number of training iterations & $N_{policy}=  80$, $N_{aux}= 6$ \\
        Deadline coefficient & $\lambda_k^{penalty}=10$ \\
        Policy update frequency & $N_{update}=4$ \\
        Entropy weight &  $\beta=1.\mathrm{e}{-4}$ \\
        \bottomrule
    \end{tabular}
    }
\end{table}

\textbf{Deployment Performance:}
Fig. \ref{fig:deployment-performance} demonstrates the performance trade-offs between perceived PSNR, task response time (RT), and energy consumption (EC) in an environment with $K=3$ users and objective weights of $w_0=0.35$, $w_1=0.85$, $w_2=0.15$.
\bb{PSNR measures the perceived viewport PNSR for the elastic tasks that have met their deadline.
} 
The weights are chosen from the Pareto frontier set presented in Fig.~\ref{fig:obj-space}.
Each point represents the average values of these metrics, with vertical and horizontal bars denoting the 5th and 95th percentiles.
Furthermore, Table \ref{tab:obj-values} reports the numerical values of these points along with deadline violation (DV) percentage.
The results show that both CPPG and IPPG achieve near-optimal performance, with CPPG performing slightly better.
Specifically, CPPG improves the overall performance (average reward) by 5.54\% and perceived PSNR by 43.21\%, response time by 42.35\%, and energy consumption by 56.83\% compared to EA-Offloader.
This advantage arises from CPPG’s full observability of the system state \bb{and the joint action space (\ie elasticity parameter and offloading decision)}, as opposed to EA-Offloader’s fixed elasticity parameter. 
Consequently, CPPG can more effectively balance trade-offs across users \bb{and available resources.}
The improvement, however, comes at the expense of higher computational complexity due to centralized decision-making.
Since the decision-making process runs on the MEC unit, this overhead is not prohibitive.
Moreover, CPPG’s state space scales with the number of users in the environment, resulting in higher sample complexity during training and potential scalability limitations.
\begin{table}[t]
    \centering
    \caption{Average reward, video quality in terms of PSNR, response time (RT), energy consumption (EC) for $K=3$ users with weights $w_0=0.35$, $w_1=0.85$, and $w_2=0.15$. Both CPPG and IPPG achieve near-optimal performance with no deadline violations, while EGreedy and EA-Offloader show larger variability, lower rewards, and higher violation rates.}
    \label{tab:obj-values}
    \resizebox{.8\linewidth}{!}{
    \begin{tabular}{|>{\bfseries}l|c|c|c|c|c|}
        \hline
        \textbf{Baseline} & \textbf{Reward} & \textbf{PSNR (dB)} & \textbf{RT (s)} & \textbf{EC (mJoule)} & \textbf{DV \%} \\
        \hline
            Optimal & $16.48$ &$49.24 \pm 1.03$ &$0.57 \pm 0.07$ & $1.80 \pm 0.98$ &$\mathbf{0.0}$ \\
            CPPG & $\mathbf{16.36}$ &$\mathbf{48.35 \pm 0.80}$ &$\mathbf{0.49 \pm 0.13}$ & $1.01 \pm 0.74$ &$\mathbf{0.0}$ \\
            IPPG & $16.30$ &$48.14 \pm 0.57$ &$0.53 \pm 0.03$ & $\mathbf{0.61 \pm 0.04}$ &$\mathbf{0.0}$ \\
            EGreedy & $13.47$ &$33.81 \pm 15.68$ &$0.85 \pm 0.73$ & $2.06 \pm 2.39$ &$40.7$ \\
            EA-Offloader & $15.50$ &$38.63 \pm 16.44$ &$0.85 \pm 0.35$ & $2.34 \pm 0.92$ &$32.4$ \\
        \hline
    \end{tabular}
    }
\end{table}

\begin{figure*}
    \centering
    \begin{subfigure}{0.48\linewidth}
        \centering
        \includegraphics[width=\columnwidth]{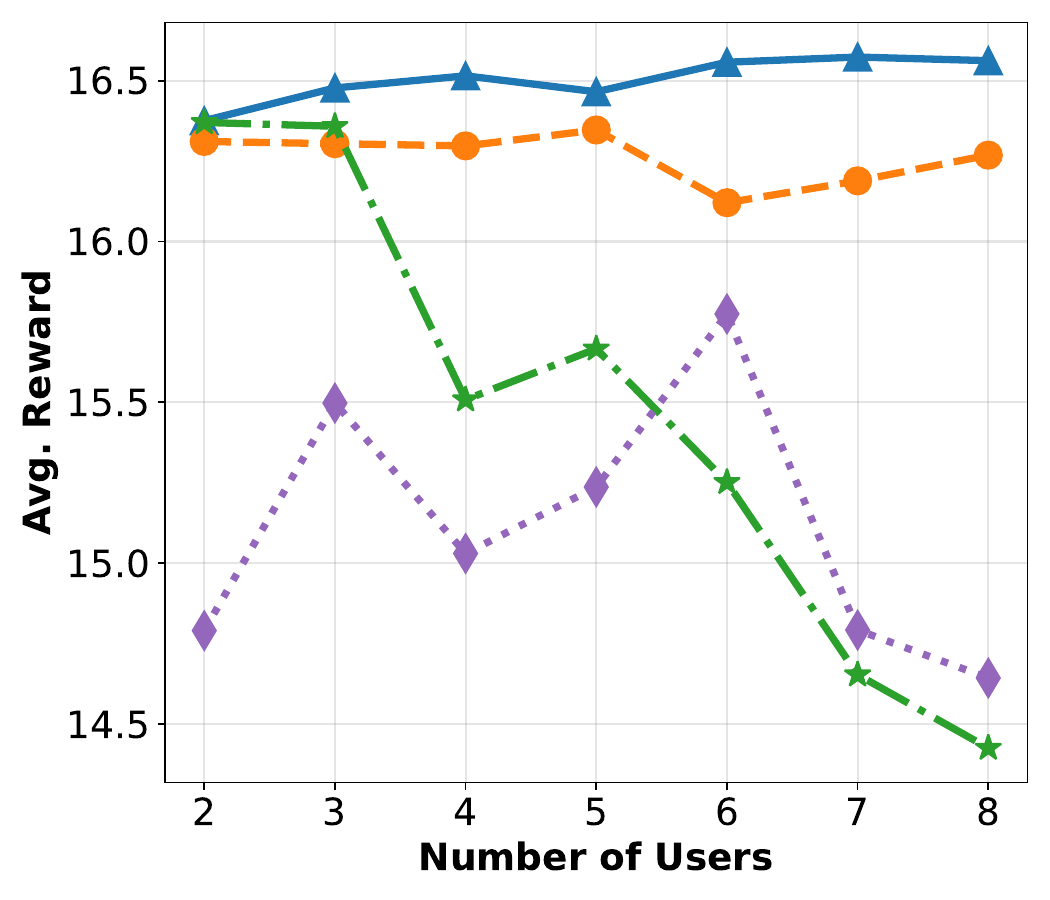}
        \subcaption{}
        \label{fig:scalability-a}
    \end{subfigure}
    \begin{subfigure}{0.48\linewidth}
    \centering
        \includegraphics[width=\columnwidth]{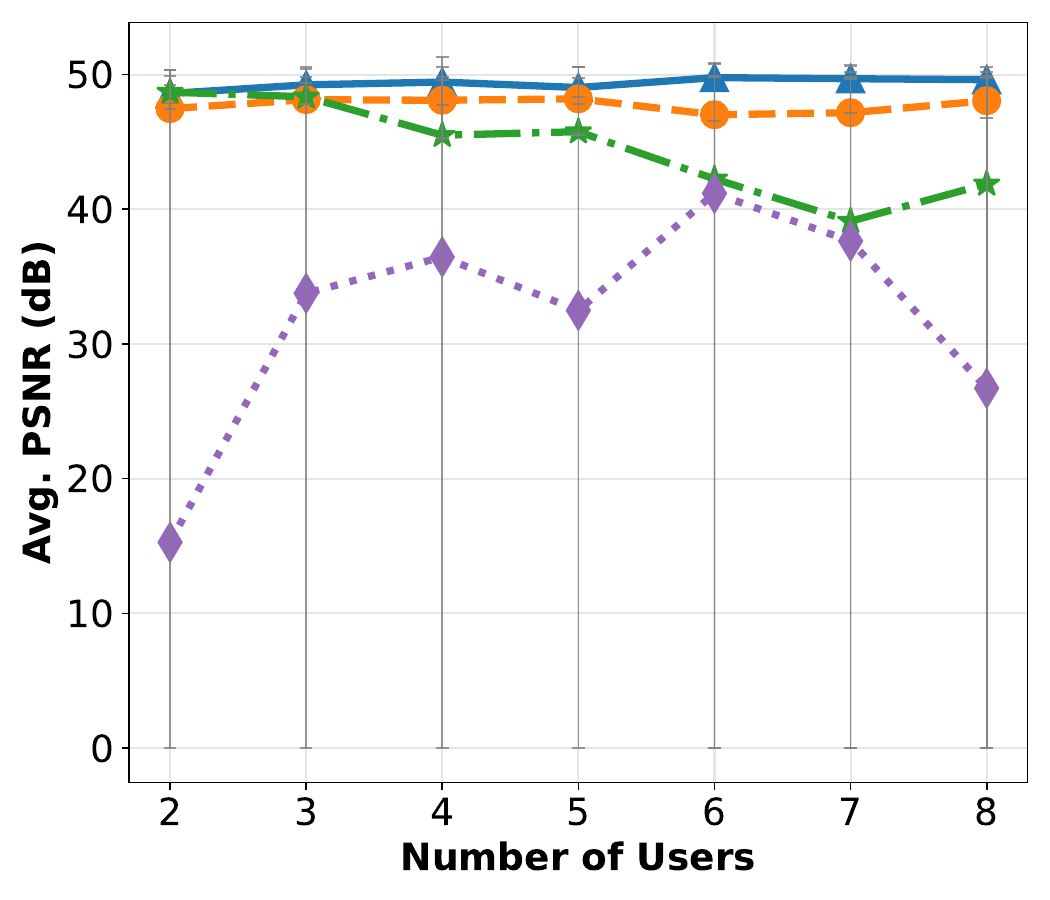}
        \subcaption{}
        \label{fig:scalability-b}
    \end{subfigure}
    \begin{subfigure}{0.48\linewidth}
    \centering
        \includegraphics[width=\columnwidth]{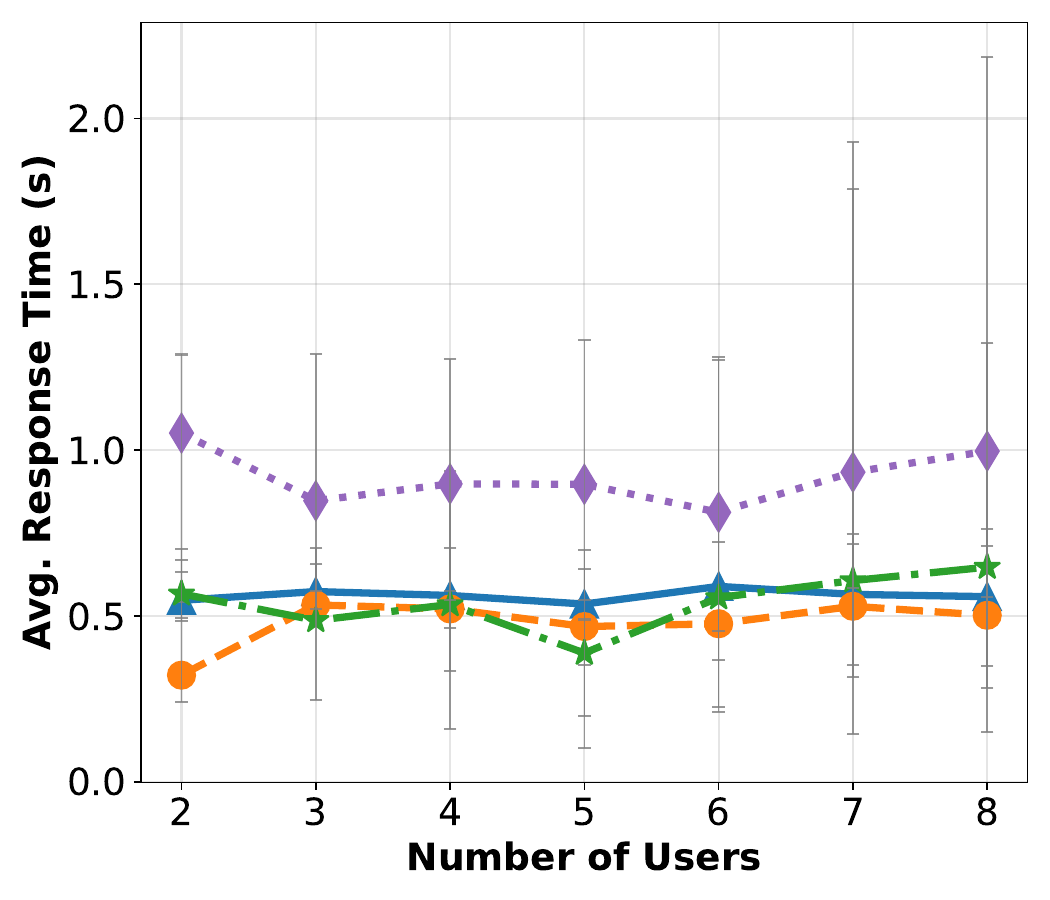}
        \subcaption{}
        \label{fig:scalability-c}
    \end{subfigure}
    \begin{subfigure}{0.48\linewidth}
    \centering
        \includegraphics[width=\columnwidth]{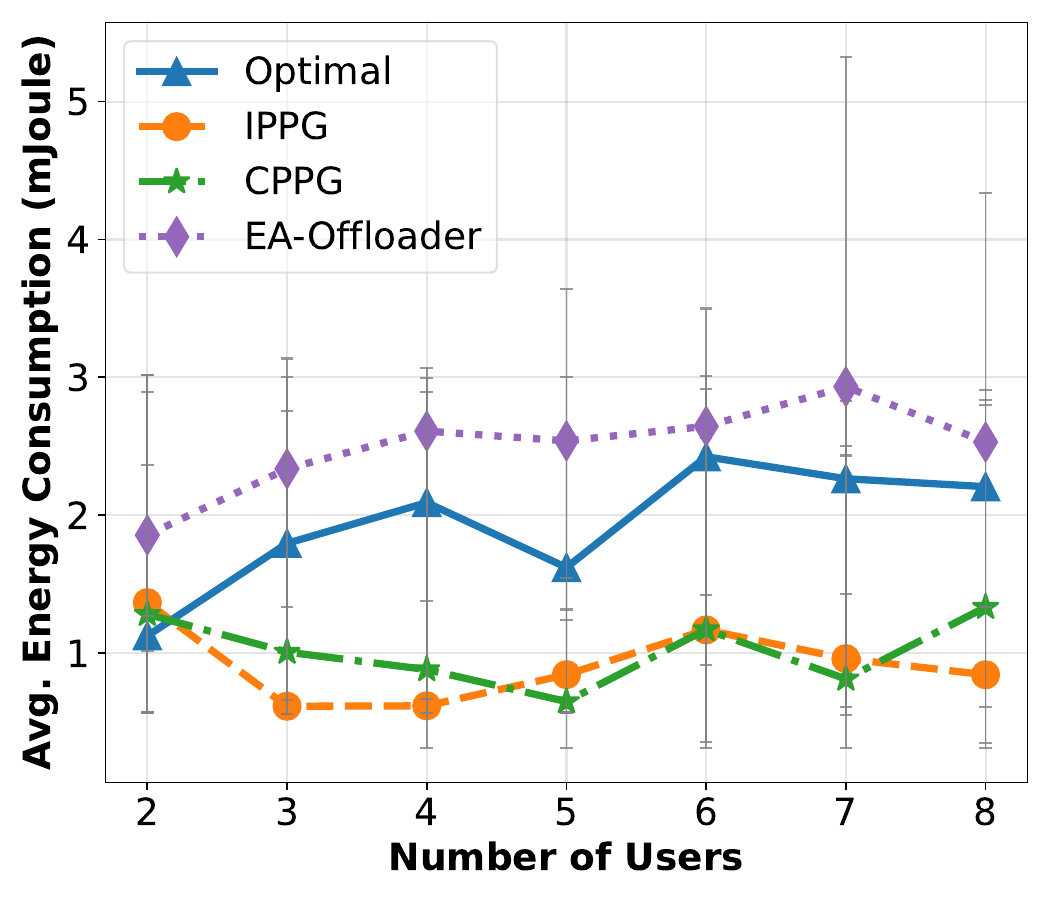}
        \subcaption{}
        \label{fig:scalability-d}
    \end{subfigure}
    \caption{
   Scalability analysis of the proposed frameworks compared to baselines as the number of users increases. Subplots show (a) average reward, (b) average perceived PSNR, (c) average response time, and (d) average energy consumption. The results indicate that IPPG maintains near-optimal performance across different user scales, while CPPG and EA-Offloader exhibit higher variability and degradation in performance.
    }
    \label{fig:scalability}
\end{figure*}

\textbf{Scalability Analysis:}
Fig.~\ref{fig:scalability} shows the performance of our proposed methods compared to the baselines as the number of users in the environment increases.
The results show that IPPG exhibits superior scalability. 
While CPPG performance degrades as the number of users increases, IPPG consistently maintains near-optimal solutions.
This improvement can be attributed to two factors: (i) IPPG has a smaller state space, yielding higher sample efficiency; and (ii) parameter sharing during training further enhances this efficiency. 
For fairness, the number of users is the only variable changed in this experiment, while all other system parameters remain fixed.
It is worth noting, however, that the objective weights selected for this experiment (Eq.~\ref{eq:qte}) do not necessarily lie on the Pareto-optimal frontier for different number of users.
Although the EGreedy baseline was tested, its performance was consistently inferior and is therefore omitted from this analysis.
Overall, the results demonstrate that IPPG achieves scalability advantages over both traditional baselines and centralized learning approaches.


\textbf{Effect of Elasticity Parameter:}
Table~\ref{tab:elasticity-analysis} evaluates the impact of the elasticity parameter on system performance under the IPPG policy with $K=5$ heterogeneous users. When the elasticity parameter is fixed to specific values ($e_k=7,\dots,1$), the performance varies significantly across metrics. High elasticity values (e.g., $e_k=7$ or $e_k=6$) lead to poor outcomes, with extremely negative rewards, zero PSNR, and very high response times and energy consumption. As elasticity decreases, the results improve, with $e_k=3$ and $e_k=2$ producing near-optimal trade-offs across reward, PSNR, response time, and energy. Notably, the fully adaptive case, where both elasticity and offloading are learned jointly by IPPG (row “Elastic”), achieves the highest overall reward ($14.74$), along with balanced quality and efficiency metrics. This demonstrates that rigidly fixing the elasticity parameter can severely bias system performance, whereas enabling IPPG to dynamically adjust elasticity allows it to generalize better across diverse workloads, maintain video quality, and optimize resource usage simultaneously.

\begin{table}[]
    \centering
    \caption{Performance of IPPG with fixed elasticity parameters ($e_k=1,\dots,7$) compared to the adaptive case (“Elastic”) where both elasticity and offloading are made by IPPG. Results are reported for $K=5$ heterogeneous users streaming different videos. The adaptive policy achieves the highest reward and balanced trade-offs across PSNR, response time, and energy consumption, highlighting the advantage of elasticity adaptation over fixed settings.}
    \label{tab:elasticity-analysis}
    \resizebox{.8\linewidth}{!}{
    \begin{tabular}{|>{\bfseries}c|c|c|c|c|c|}
    \hline
    \textbf{Method} & \textbf{Reward} &  \textbf{PSNR (dB)} & \textbf{RT (s)} & \textbf{EC (mJoule)} & \textbf{DV (\%)} \\
    \hline
        $u_k = 7$ & $-72.61$ &$15.00 \pm 0.00$ &$8.99 \pm 2.72$ & $35.88 \pm 10.18$ &$100.00$ \\
$u_k = 6$ & $-15.59$ &$15.00 \pm 0.00$ &$3.90 \pm 1.45$ & $15.58 \pm 6.09$ &$100.00$ \\
$u_k = 5$ & $6.77$ &$19.11 \pm 11.67$ &$1.87 \pm 1.03$ & $6.76 \pm 4.18$ &$88.89$ \\
$u_k = 4$ & $12.31$ &$33.15 \pm 17.42$ &$1.19 \pm 1.11$ & $2.61 \pm 4.32$ &$47.78$ \\
$u_k = 3$ & $14.59$ &$42.67 \pm 11.70$ &$0.78 \pm 0.49$ & $1.21 \pm 1.15$ &$15.00$ \\
$u_k = 2$ & $14.53$ &$42.53 \pm 9.31$ &$\mathbf{0.56 \pm 0.43}$ & $\mathbf{1.14 \pm 1.49}$ &$\mathbf{10.00}$ \\
$u_k = 1$ & $13.68$ &$39.60 \pm 10.33$ &$0.56 \pm 0.68$ & $\mathbf{0.97 \pm 2.74}$ &$14.44$ \\
Elastic & $\mathbf{14.74}$ &$\mathbf{42.98 \pm 10.15}$ &$0.61 \pm 0.41$ & $1.36 \pm 1.44$ &$11.11$ \\
    \hline
    \end{tabular}
    }
\end{table}

\begin{table}[]
    \centering
    \caption{
    Performance of IPPG and CPPG under heterogeneous content with $K=8$ users simultaneously streaming videos of varying spatio-temporal complexity. Results highlight that static content (e.g., Bridge, StudyRoom) yields consistently high PSNR and low latency, while motion-rich content (e.g., Basketball) remains more challenging. On average, IPPG achieves +12.57\% reward, +19.54\% PSNR, +20\% latency reduction, and +35.13\% energy savings compared to CPPG. In addition to higher mean performance, IPPG exhibits lower variance across heterogeneous users, ensuring a more uniform QoE distribution.
    }
    \label{tab:video-analysis}
    \resizebox{\columnwidth}{!}{
    \begin{tabular}{>{\bfseries}l|c|c|c|c|c|c|c|c|c|c}
    \multicolumn{1}{c|}{} & \multicolumn{5}{c|}{\textbf{IPPG}} & \multicolumn{5}{c}{\textbf{CPPG}} \\
    \hline
\textbf{Video Name} & \textbf{Reward}  & \textbf{PSNR (dB)} & \textbf{RT (s)} & \textbf{EC (mJoule)} & \textbf{DV (\%)} & \textbf{Reward} & \textbf{PSNR (dB)} & \textbf{RT (s)} & \textbf{EC (mJoule)} & \textbf{DV (\%)} \\
    \hline
        Academic & $16.21$ & $48.16 \pm 1.82$ & $0.63 \pm 0.15$ & $0.77 \pm 0.23$ & $0.00$ & $15.52$ & $43.50 \pm 9.21$ & $0.58 \pm 0.23$ & $1.24 \pm 1.32$ & $\mblue{8.33}$ \\
        Basketball & $10.18$ & $\mgreen{27.58 \pm 13.53}$ & $\mgreen{1.24 \pm 0.55}$ & $3.08 \pm 2.13$ & $\mgreen{52.78}$ & $10.41$ & $\mgreen{22.07 \pm 13.52}$ & $\mgreen{1.33 \pm 0.63}$ &$5.71 \pm 2.83$ & $\mblue{77.78}$ \\
        Bridge & $16.10$ & $\mathbf{47.13 \pm 1.15}$ & $\mathbf{0.36 \pm 0.13}$ & $0.62 \pm 0.69$ & $0.00$ & $13.95$ & $44.21 \pm 8.95$ & $0.53 \pm 1.21$ & $0.64 \pm 0.11$ &$\mblue{8.33}$ \\
        GateNight & $16.27$ & $48.02 \pm 1.76$ & $0.40 \pm 0.15$ & $1.29 \pm 1.05$ & $0.00$ & $14.80$ & $\mblue{16.94 \pm 8.09}$ & $\mblue{1.15 \pm 0.07}$ & $1.32 \pm 0.08$ & $\mblue{94.44}$ \\
        Runner & $15.53$ & $46.44 \pm 1.69$ & $0.60 \pm 0.12$ & $1.45 \pm 1.20$ & $0.00$ & $14.50$ & $42.84 \pm 1.30$ & $0.49 \pm 0.02$ & $0.56 \pm 0.03$ & $0.00$ \\
        SiyuanGate & $17.32$ & $51.00 \pm 1.70$ & $0.52 \pm 0.08$ & $0.60 \pm 0.09$ & $0.00$ & $12.48$ & $45.49 \pm 15.30$ & $\mblue{1.14 \pm 1.64}$ & $1.97 \pm 0.48$ & $\mblue{19.44}$ \\
        SouthGate & $15.58$ & $42.57 \pm 11.48$ & $0.85 \pm 0.17$ & $1.13 \pm 0.45$ & $\mgreen{13.89}$ & $13.71$ & $40.34 \pm 6.54$ & $0.68 \pm 0.21$ & $1.22 \pm 1.37$ & $\mblue{5.56}$ \\
        StudyRoom & $16.71$ & $\mathbf{49.31 \pm 2.20}$ & $\mathbf{0.53 \pm 0.11}$ & $0.63 \pm 0.16$ & $0.00$ & $14.69$ & $45.92 \pm 9.47$ & $0.54 \pm 0.82$ & $2.16 \pm 3.71$ & $\mblue{8.33}$ \\
        \hline
        Average &$15.49$ &$\mred{45.02 \pm 9.46}$ &$\mred{0.64 \pm 0.35}$ &$1.20 \pm 1.25$ &$0.08$ &$13.76 $ &$\mred{37.66 \pm 14.49}$ &$0.80 \pm 0.87$ &$1.85 \pm 2.35$ &$0.28$ \\
    \end{tabular}
    }
\end{table}

\textbf{Heterogeneous Content Analysis:}
To further assess robustness, we evaluate the performance of IPPG in a more stochastic and uncertain environment where $K=8$ users simultaneously stream videos with distinct spatio-temporal characteristics (Table \ref{tab:video-analysis}). The results demonstrate that IPPG sustains strong performance across diverse workloads, achieving a 12.57\% improvement in overall reward, 19.54\% in PSNR, 20\% improvement in response time, and 35.13\% improvement in energy consumption compared to CPPG.
A closer inspection of individual content types, however, reveals important differences.
Static or slowly varying content such as Bridge and StudyRoom (See bold values in Table \ref{tab:video-analysis}) yields consistently high PSNR ($\approx$47–49 dB) and low latency ($\approx$0.4–0.5s) under IPPG.
In contrast, highly dynamic videos such as Basketball present a much greater challenge: although IPPG still improves over CPPG, its PSNR shows wide fluctuations and latency remains relatively high.
This contrast highlights the difficulty of supporting motion-rich VR scenarios and underscores the importance of adaptive methods.
Another key advantage of IPPG is its stability.
Across most videos, IPPG produces smaller standard deviations in PSNR, response time, and energy consumption than CPPG, which often exhibits large variability (\eg GateNight and Basketball).
This reduced variance translates into a more predictable and reliable user experience.
Taken together, these findings show that IPPG generalizes effectively across heterogeneous video content, delivering both higher average performance and more consistent per-user experiences.
From a practical standpoint, this adaptability is crucial: future VR/AR platforms must seamlessly handle both static workloads (\eg lectures or meetings) and dynamic workloads (\eg live sports or gaming).
The ability of IPPG to maintain near-optimal trade-offs across such extremes reinforces its potential for real-world deployment.

\section{Conclusion}\label{sec:conclusion}
In this paper, we studied the problem of elastic task offloading in multi-user, multi-connectivity immersive computing systems.
We formulated a constrained QoE and energy optimization problem that explicitly captures the trade-offs between video quality, task response time, and energy consumption under heterogeneous communication and computation resources.
To address the challenges posed by time-varying \bb{network conditions} and partial observability, we developed the ElasticVR framework consisting of two MARL approaches: the CPPG and the IPPG.
Through extensive simulations with real-world 4G, 5G, and WiGig traces and diverse $360^\circ$ video workloads, we demonstrated that both CPPG and IPPG achieve near-optimal performance while satisfying strict task deadlines.
CPPG benefits from full system observability and achieves stronger global coordination across users, but at the cost of higher training complexity and limited scalability.
In contrast, IPPG offers a scalable and efficient alternative by leveraging parameter sharing and decentralized execution, maintaining strong performance across increasing numbers of users and heterogeneous video scenarios.
Our analysis further showed the importance of elasticity adaptation: rigid baselines that fix elasticity consistently underperformed, whereas dynamic elasticity control enables the ElasticVR framework to sustain balanced trade-offs across QoE, latency, and energy consumption.
\bb{
Moreover, our analysis shows that CPPG outperforms IPPG (\ie reduces latency by 28\% and energy consumption by 78\%) in environments with smaller number of users, while IPPG outperforms CPPG (\ie 19.54\% improvement in PSNR, 20\% reduction in response time, and 35.13\% reduction in energy consumption) as the number of users in an environment increases.
}
Overall, this work highlights the effectiveness of MARL-based approaches for elastic task offloading in next-generation immersive systems, and provides insights into the fundamental trade-offs between centralized coordination and decentralized scalability.
In future work, we plan to extend our framework to support heterogeneous workloads, incorporate user-level fairness, compare against a broader set of state-of-the-art reference methods, and 
explore the elastic edge computing methodology in other emerging applications in next-generation wireless networks.

\section{Acknowledgment}
This work was supported in part by NSF grants 1955561, 2212565, 2323189, 2434113, 2514415, 2032033, 2106150, and 2346528. The work of Jacob Chakareski was additionally supported in part by NIH award R01EY030470; and by the Panasonic Chair of Sustainability at NJIT.

{
\bibliographystyle{ACM-Reference-Format}
\bibliography{ref2}
}
\end{document}